\documentclass[a4paper,11pt]{article}
\pdfoutput=1 

\usepackage{jcappub} 

\usepackage[T1]{fontenc} 

\DeclareMathOperator{\sech}{sech}

\usepackage{comment}
\usepackage{ulem}
\usepackage{booktabs}
\usepackage{enumerate}   
\usepackage{orcidlink}
\newcommand\scalemath[2]{\scalebox{#1}{\mbox{\ensuremath{\displaystyle #2}}}}

\title{\huge\boldmath 
Shadows, rings and optical appearance of a magnetically charged regular black hole illuminated by various accretion disks
}

\author[a]{\large Soroush Zare,\orcidlink{0000-0003-0748-3386}}
\author[a,1]{Luis M. Nieto,\orcidlink{0000-0002-2849-2647}\note{Corresponding author.}}
\author[b]{Xing-Hui Feng,\orcidlink{0000-0003-3486-7828}}
\author[c,d]{ Shi-Hai Dong,\orcidlink{0000-0002-0769-635X}}
\author[e,f]{and Hassan Hassanabadi, \orcidlink{0000-0001-7487-6898}}

\affiliation[a]{ Departamento de F\'{\i}sica Te\'orica, At\'omica y Optica and Laboratory for Disruptive \\ Interdisciplinary Science (LaDIS), Universidad de Valladolid, 47011 Valladolid, Spain}
\affiliation[b]{Center for Joint Quantum Studies and Department of Physics, School of Science,
	Tianjin University, Tianjin 300350, China}
\affiliation[c]{Research Center for Quantum Physics, Huzhou University, Huzhou, 313000, PR China}
\affiliation[d]{Centro de Investigación en Computación, Instituto Politécnico Nacional, UPALM, CDMX 07700, Mexico}

\affiliation[e]{Faculty of Physics, Shahrood University of Technology, Shahrood, Iran}
\affiliation[f]{Department   of   Physics,   University   of   Hradec   Kr\'{a}lov\'{e}, Rokitansk\'{e}ho   62,   500   03   Hradec   Kr\'{a}lov\'{e},   Czechia}

\emailAdd{szare@uva.es}
\emailAdd{luismiguel.nieto.calzada@uva.es} 
\emailAdd{xhfeng@tju.edu.cn}
\emailAdd{dongsh2@yahoo.com}
\emailAdd{h.hasanabadi@shahroodut.ac.ir}

 \abstract{The Event Horizon Telescope (EHT) imaging of the supermassive black holes at the centers of Messier 87 galaxy (M87) and the Milky Way galaxy (Sgr A) marks a significant step in observing the photon rings and central brightness depression that define the optical appearance of black holes with an accretion disk scenario.
Inspired by this, we take into account a static and spherically symmetric magnetically charged regular black hole (MCRBH) metric characterized by its mass and an additional parameter $q$, which arises from the coupling of Einstein gravity and nonlinear electrodynamics (NLED) in the weak field approximation.	
This parameterized model offers a robust foundation for testing the coupling of Einstein gravity and NLED in the weak-field approximation, using the EHT observational results.
In this study, we investigate the geodesic motion of particles around the solution, followed by a discussion of its fundamental geometrical characteristics such as scalar invariants. 
Using null geodesics, we examine how the model parameter influences the behavior of the photon sphere radius and the associated shadow silhouette.  We seek constraints on $q$ by applying the EHT results for supermassive black holes M87* and Sgr A*. 
Furthermore, it is observed that the geodesics of time-like particles are susceptible to variations in $q$, which can have an impact on the traits of the innermost stable circular orbit and the marginally bounded orbit. 
Our primary objective is to probe how the free parameter $q$ affects various aspects of the accretion disk surrounding the MCRBH using the thin-disk approximation. 
Next, we discuss the physical characteristics of the thin accretion disk as well as the observed shadows and rings of the MCRBH, along with its luminosity, across various accretion models.
Ultimately, variations in accretion models and the parameter $q$ yield distinct shadow images and optical appearances of the MCRBH.
}


\makeatletter
\gdef\@fpheader{}
\makeatother

\begin{document}
\maketitle
\flushbottom

\section{Introduction}
Credible alternative theories of gravity must adhere to established observational constraints. Specifically, they must satisfy rigorous solar system constraints for weak gravitational fields \cite{WillLRR2006,WillCUP2018}.  However, observations in the vicinity of compact objects such as black holes (BHs) or neutron stars, which have strong gravitational fields, provide a wide range of possibilities \cite{BertiCQG2015,BarackCQG2019,Saridakis2021,HuangJCAP2024}. 
BHs, exceptionally compact celestial entities, were initially theorized by Karl Schwarzschild in 1916 \cite{Schwarzschild} following the inception of General Relativity (GR), sparking widespread interest from the perspective of observations.

The defining characteristic of a BH is its event horizon. This is the boundary beyond which particles cannot escape the BH's gravitational pull as they approach to interior. Any particle, including light, that crosses this threshold is trapped. However, outside the event horizon, even beyond its edge, light can escape the gravitational pull and travel to infinity.
\cite{SyngeMNRAS1966}.
Although the presence of BHs is undeniable, detecting them remains immensely challenging.  The concept of the BH shadow is crucial in this context, as the act of visualizing a BH inspires an examination of the exceptionally extreme gravity regime at its horizon.
The recent detection of gravitational-wave (GW) signals emitted from binary BH mergers using the Laser-Interferometer Gravitational Wave-Observatory (LIGO) experiments, as well as the discovery of a wide star-BH binary system through radial-velocity measurements, provide strong confirmation of the presence of BHs in the universe \cite{AbbottPRL2016,AbbottPRD2020,AbbottPRL2020,LiuNature20219}.
The ultra-high angular resolution images of M87* and Sgr A*, released by the EHT Collaboration \cite{AkiyamaL12019,AkiyamaL52019,AkiyamaL62019,AkiyamaL122022,AkiyamaL172022}, provide more obvious proof of the presence of BHs \cite{DoScience2019,GravityCollaborationAA2022}. 
The images contain a dim area at the heart known as the BH shadow \cite{LuminetAA1979,FalckeApJL13,NarayanApJL2019,PerlickPR2022,FengEPJC2020, HuEPJC2022,WenePJC2023}, surrounded by bright rings corresponding to the photon sphere. The phenomenon of light ray deflection near a BH, known as lensing effect \cite{EinsteinS1936,VirbhadraPRD2000,PerlickLRR2004,CunhaGRG2018}, is widely recognized. 
Indeed, photon trajectories emitted by a far-off light source behind the BH are bent by its gravitational pull, forming a BH shadow and surrounding photon sphere.
This leads to a decrease in the observed specific intensity within a well-defined border, creating a zone of reduced brightness on the remote image plane. 
After the publication of these images, numerous studies flooded the corresponding literature, with many concentrating on investigating how the BH shadow could improve our comprehension of their characteristics \cite{TsupkoPRD2017,BroderickApJ2022,MizunoNatAst2018}
and provide insights into deviations in the spacetime geometry \cite{AtamurotovPRD2013,AbdujabbarovSS2016,AbdikamalovPRD2019,AtamurotovPRD2015,AtamurotovCPC2023,BelhajPLB2021,BelhajCQG2021,CunhaPLB2017,WeiJCAP2019,LingPRD2021,TsukamotoPRD2018,AraujoFilhoCQG2024,RayimbaevPoDU2022,PerlickPRD2018,VagnozziCQG2023,KocherlakotaPRD2021}.
Such deviations could stem from parameters in different alternative theories of gravity \cite{UniyalPoDU2023,Uniyal2023,LambiaseEPJC2023,KhodadiPRD2022,KhodadiJCAP2021,PanotopoulosPRD2021,EslamPanahEPJC2024}, or from the astrophysical surroundings in which the BH is situated \cite{WuPoDU2024,PantigJCAP2022,XuJCAP2018,CapozzielloJCAP2023,Capozziello2023,KonoplyaPLB2019,KonoplyaApJ2022}.

Recent astrophysical measurements indicate that emissions from BHs primarily originate from disk-shaped accretions \cite{RemillardARAA2006,YuanARAA2014}. 
Studying accretion disks around BHs offers a potential avenue for distinguishing between GR and alternative theories. These disks consist of diffuse material orbiting a central compact object \cite{HeEPJC2022}.
The steady-state thin accretion disk represents the most basic theoretical model, where the disk is considered to have minimal thickness \cite{ShakuraAA1973,Novikov1973,PageApJ1974}. Extensive discussions on the physical properties of matter composing a thin accretion disk in various background spacetimes can be found in the literature, such as in \cite{BambiApJ2011,BambiApJ2012,ChenPLB2011,PerezAA2013, CollodelApJ2021,BoshkayevPRD2021,HarkoPRD2009,LiuJCAP2022,UniyalPoDU2023,Uniyal2023,FathiEPJC2023}.
The radiation emanating from the accretion disk hinges on the particle motion within the gas and the spacetime configuration surrounding the BH. The signatures observed in the energy flux and spectrum radiated by the disk offer insights into BHs and serve as a means to test modified gravity theories \cite{BambiRMP2017}.
Hence, one of the objectives of our study is to comprehend how a magnetically charged BH influences the trajectories of particles orbiting around it along the geodesics path.

The luminous area surrounding the dark region emanates from the accretion material encircling real astrophysical BHs, significantly influencing the observed shape of the BH due to variations in accretion material distributions. While replicating a realistic accretion disk through theoretical means poses challenges, the first image of a BH with a thin accretion disk was analytically computed in \cite{LuminetAA1979}. This calculation revealed the presence of primary and secondary images outside the BH shadow.
Subsequently, in \cite{BambiPRD2013}, it was noted that distinguishing between a Schwarzschild BH and a static wormhole based on shadow images is relatively feasible.
Recent studies on Schwarzschild BHs with both thin and thick accretion disks have revealed that the lensed ring, along with the photon ring, imparts significant observed brightness to the BH image. 
The number of times photons intersect with the accretion disk is a crucial factor in determining the optical characteristics of shadows. The photon ring comprises light rays intersecting the accretion disk at least three times \cite{GrallaPRD2019}.
Additionally, spherical accretion has been employed as another form of accretion to examine the Schwarzschild BH image \cite{FalckeApJL13,NarayanApJL2019}.
According to the literature, the shadow is a solid attribute whose size and shape are Mainly affected by the spacetime geometry rather than characteristics of the accretion disk.
Recent studies on photon rings and the observational characteristics of BHs have attracted significant attention, revealing images of static BHs with accretion models beyond GR \cite{BromleyApJ1997,BambiPRD2013,JaroszynskiAA1997,GrallaPRD2019,ZengEPJC2020,PengCPC2021,ZengPRD2023,GuoPRD2022,WangPRD2023,WaliaJCAP2023,FathiEPJC2023,LiEPJC2021,LambiaseJCAP2023,ZengEPJC2020-2,HuEPJC2022,ShaikhMNRAS2019}.

While it has achieved significant successes, GR also faces challenges, particularly at the core of standard BH solutions. According to GR, singularities (points where the laws of physics cease to apply) are predicted, calling into question the credibility of Einstein’s theory.
One potential approach to address these issues is to consider suitable matter distributions, which can lead to singularity-free BH solutions within the framework of GR \cite{dePaulaPRD2023}.
James Bardeen introduced the first metric tensor for a nonsingular BH geometry in 1968 \cite{Bardeen}. Through coupling GR with nonlinear electrodynamics (NLED), it has been demonstrated that various exact charged regular BH solutions can be derived \cite{DymnikovaGRG1992,AyonBeatoPRL1998,AyonBeatoGRG1999,AyonBeatoPLB1999}. Within these models, the Bardeen geometry can be described as a regular BH caused by a nonlinear magnetic \cite{AyonBeatoPLB2000} or electric monopole \cite{RodriguesJCAP2018}.
One of the earliest covariant models of NLED was introduced in 1934, known as Born-Infeld electrodynamics, aiming to achieve a finite self-energy density for the electric charge \cite{BornPRSA1934-1,BornPRSA1934-2}.
One significant outcome of employing NLED in BH physics is the interpretation of photon motion as a null geodesic within an effective geometry \cite{NovelloPRD2000,HabibinaPRD2020,AmaroPRD2020}, distinct from the spacetime geometry itself.
Photons in linear electrodynamics adhere to the null geodesics of the standard geometry. Thus, photons and massless particles have identical equations of motion. However, in NLED theory, photons are considered to travel along the null geodesics of an effective geometry, which deviates from the standard geometry \cite{StuchlikEPJC2019}.
It is observed that when Maxwell's weak field limit is met, as is the case with the Ay\'{o}n-Beato and Garc\'{i}a solution, the effective geometry simplifies to the standard geometry in the weak field limit \cite{AyonBeatoPLB1999,dePaulaPRD2023,MatyjasekPRD2004,MyungPRD2007,MyungPLB2008,UniyalPoDU2023,Uniyal2023,OkyayJCAP2022}. 

Inspired by the aforementioned literature, we aim to study how the parameter $q$ affects circular orbits, and the shadow cast using the null geodesics method. We will also examine various aspects of the accretion disk around the MCRBH using the thin-disk approximation. To achieve this, we investigate the impact of $q$ on the thin accretion disk's physical properties via the Novikov-Thorne model, as well as the observed shadow and optical appearance of the MCRBH with both thin disk-shaped and spherical accretion models.
The structure of this paper is as follows.
In Section~\ref{Sec2}, the spacetime of an MCRBH and the associated scalar invariants are explored.
In Section~\ref{Sec3}, the impacts of the model parameter $q$ on certain BH characteristics, such as circular orbits corresponding to null geodesics and time-like geodesics, and the BH shadow, are investigated. Furthermore, constraints on $q$ are imposed based on the bounds inferred by the EHT on the Schwarzschild shadow radius of M87* and Sgr A*.
In Section~\ref{Sec4}, the physical characteristics of the thin accretion disk surrounding the MCRBH are examined. Subsequently, an optically thin accretion disk model is investigated, depicting the shadow contour, rings, and corresponding observed luminosity for a distant observer. Following this, exploration is conducted into the shadow image and luminosity in spherical static and infalling accretion formalism. Finally, discussions and conclusions are presented in Section~\ref{SecC}.

\section{Magnetically charged regular black hole}\label{Sec2}
\begin{figure}[htb]
\centering 
\includegraphics[width=.49\textwidth]{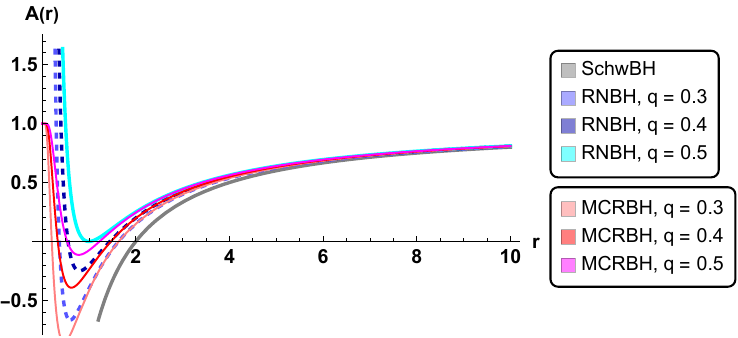}
	\hfill
\includegraphics[width=.49\textwidth]{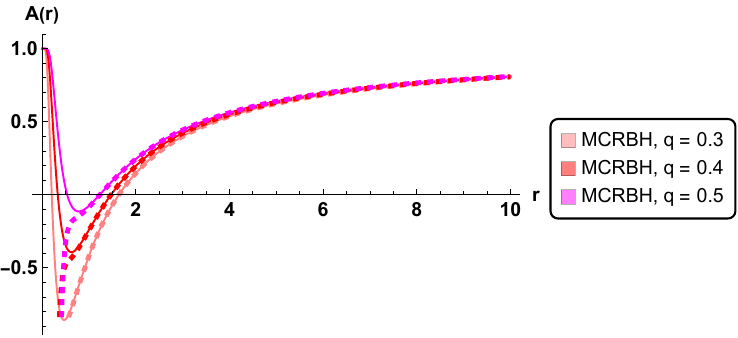}
	\caption{\label{fig1-1} Plot of the metric function $\text{A}(r)$ as it varies with different values of $q$.  
	It is clear that the MCRBH solution resembles the RN solution closely, especially when the value of the charge is smaller, except in the vicinity of the singular, that is, $r \rightarrow 0$. However, as $r \rightarrow \infty$, the MCRBH solution converges to the RN solution.	
	The left panel illustrates a comparison between RNBH and MCRBH solutions. The RNBH solution is depicted with both dashed and solid blue lines.
	The right panel displays the approximate form of the MCRBH solution, indicated by dashed pink, red, and magenta lines.}
\end{figure}

Here, we will provide a brief review of a MCRBH  \cite{AyonBeatoPLB1999,MatyjasekPRD2004,MyungPRD2007,MyungPLB2008}. We begin with a minimally coupled four-dimensional gravity action to NLED, which is given by
\begin{equation}\label{Action}
I=\frac{1}{16\pi}\int d^{4}x \sqrt{-g}\, \left[R-\mathcal{L}_{\text{M}}\left(B\right)\right],
\end{equation}
where $R$ stands for the Ricci scalar and the Lagrangian for matter, $\mathcal{L}_{\text{M}}\left(B\right)$, is a functional of $B=F_{\mu\nu}F^{\mu\nu}$ with the electromagnetic field $F_{\mu\nu}=\partial_{\mu}A_{\nu}-\partial_{\nu}A_{\mu}$, that is defined as follows \cite{MatyjasekPRD2004,MyungPRD2007,MyungPLB2008}
\begin{equation}\label{LagMatter}
\mathcal{L}_{\text{M}}\left(B\right) = B \cosh^{-2}\left[a\left(\frac{B}{2}\right)^{1/4}\right],
\end{equation}
wherein the adjustment of the free parameter $a$ will provide regularity at the center point. The Einstein–Maxwell theory can be recovered in favor of the Reissner–Nordstr\"{o}m BH (RNBH) in the limit $a\rightarrow 0$. 
Given the variation of the action \eqref{Action}, matter \eqref{LagMatter} and the tensor field $F_{\mu\nu}$, the following equations of motion can be obtained
\begin{equation}
\nabla_{\mu}\left(\mathcal{L}_{B}F^{\mu\nu}\right)=0, \quad \nabla_{\mu}\left({}^{\ast}F^{\mu\nu}\right)=0,
\end{equation}
in which $\mathcal{L}_{B}=\frac{d\mathcal{L}_{\text{M}}(B)}{dB}$ and the asterisk symbol represents the Hodge duality. 
The only non-zero components of $F_{\mu\nu}$ in the spherically symmetric case are a radial 
magnetic field $F_{\theta\varphi} = -F_{\varphi\theta} = Q \sin\theta$ and a radial electric field $F_{tr} = -F_{rt} = E(r)$ \cite{AyonBeatoPLB1999,MatyjasekPRD2004,MyungPRD2007,MyungPLB2008}. 
The Einstein equations are thus obtained by varying the action with regard to the metric tensor $g_{\mu\nu}$ \cite{Einstein1916} as
\begin{equation}\label{EinsteinEq}
R_{\mu\nu}-\frac12 g_{\mu\nu} R = 8 \pi T_{\mu\nu},
\end{equation}
using the stress-energy tensor
\begin{equation}\label{EnergyTensor}
T_{\mu\nu} = \frac{1}{4\pi} \left(\mathcal{L}_{B}F_{\tau\mu}F^{\tau}_{\nu}-\frac14 g_{\mu\nu}\mathcal{L}_{\text{M}}(B)\right).
\end{equation}
Here, $R_{\mu\nu}$ denoted the Ricci tensor. 
For a configuration that is both static and spherically symmetric, the spacetime can be characterised by the following metric \cite{AyonBeatoPLB1999}
\begin{subequations}\label{metric1}
	\begin{align}
		ds^{2} = - \text{A}(r) dt^{2} + \text{B}(r) dr^{2} + r^{2} (d\theta^{2}+\sin^{2}\theta d\varphi^{2}),
	\end{align}
being the ``lapse function''
	\begin{align}
		\text{A}(r) = \text{B}(r)^{-1} = 1-\frac{2m(r)}{r},
	\end{align}
\end{subequations}
so that the following is the mass distribution 
\begin{equation}\label{massdistribution1}
m(r) = \frac14 \int_{}^{r} \mathcal{L}_{\text{M}}(B(r'))r'^{2} \,dr'+C,
\end{equation}
with $C$ an appropriate the integration constant. By integrating Eq. \eqref{massdistribution1} with the function $\mathcal{L}_{\text{M}}(B)$ given in Eq. \eqref{LagMatter}, where $B=\frac{2Q^{2}}{r^{4}}$, the result is \cite{AyonBeatoPLB1999}
\begin{equation}\label{massdistribution2}
	m(r) = M-\frac{Q^{3/2}}{2a}\tanh\left(\frac{aQ^{1/2}}{r}\right).
\end{equation}
This mass distribution is obtained by taking into account the condition for the Arnowitt-Deser-Misner (ADM) mass $M$, that is, at infinity, $m(\infty)=M$, a constant . 
Moreover, $Q$ is a magnetic charge. The metric function $\text{A}(r)$ is therefore determined by assigning $a= \frac{Q^{3/2}}{2M}$ as follows \cite{AyonBeatoPLB1999,MatyjasekPRD2004,MyungPRD2007,MyungPLB2008}
\begin{equation}\label{MetricFuncA}
	\text{A}(r) = 1-\frac{2M}{r}\left[1-\tanh\left(\frac{q}{r}\right)\right],
\end{equation}
where $q=\frac{Q^{2}}{2M}$ is the deviation parameter related to the magnetic charge $Q$ in MCRBH spacetime.
It should be noted that in the case where $q\rightarrow 0$, the spacetime structure can return to the Schwarzschild spacetime. The metric function $\text{A}(r)$ can be approximated asymptotically as
\begin{equation}\label{MetricFuncA2}
	\text{A}(r) = 1-\frac{2M}{r}+\frac{2M q}{r^{2}}-\frac{2Mq^{3}}{3r^{4}}+\mathcal{O}\left(\frac{1}{r^{6}}\right).
\end{equation}
This differs from the  Reissner--Nordstr\"{o}m (RN) solution.
It can be observed that the MCRBH behaves asymptotically like the RNBH, given that all charged RBHs can be expanded into a corresponding polynomial. The variation in the metrics of MCRBH and RNBH is illustrated in the left panel of Figure~\ref{fig1-1}. The similarity between the MCRBH and RNBH spacetimes, excluding the region around the singular point $r=0$, is evident when considering the MCRBH with a lower $q$, which is responsible for the BH magnetic charge. Equation \eqref{MetricFuncA2} reveals that the last term is directly proportional to $r^{-4}$, which aligns with the plot of Eq. \eqref{MetricFuncA} (refer to the right panel of Figure~\ref{fig1-1}) as the value of $r$ increases. Notably, as $r \rightarrow \infty$, the metric functions given in Eq.  \eqref{MetricFuncA} and Eq.  \eqref{MetricFuncA2} simplifies to the RNBH solution. We set $M = 1$ and consider $q>0$ in all the plots for the sake of simplicity. Figure~\ref{fig1-1} displays the variation of the metric function $\text{A}(r)$ with respect to changes in $q$.

\begin{figure}[htb]
	\centering 
	\includegraphics[width=.53\textwidth]{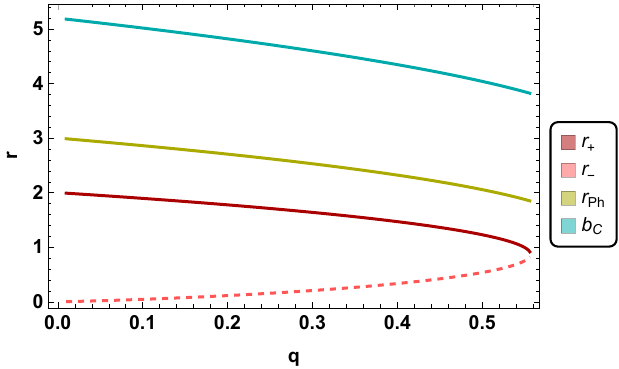}
	\caption{\label{fig1-2} The plot shows the behavior of the various radii and the critical impact parameter with varying $q$. The event horizon radius (solid red line), Cauchy horizon radius (dashed pink line), photon sphere radius (solid green line), and the critical impact parameter (solid cyan line) are all depicted.}
\end{figure}

The metric \eqref{metric1} with lapse function $\text{A}(r)$ given in Eq. \eqref{MetricFuncA} exhibits a coordinate singularity at 
\begin{equation}\label{Horizons}
\text{A}(r)=0,
\end{equation} 
allowing for up to two real positive roots, denoted as $r_{\pm}$. Here, $r_{+}$ denotes the BH event (outer) horizon, while $r_{-}$ is identified as the BH Cauchy (inner) horizon. Solving Eq. \eqref{Horizons} numerically provides the horizons for different BH parameters $q$, with a crucial threshold for the existence of the event horizon observed for $q<0.56$. Employing numerical plotting on Eq. \eqref{Horizons}, Figure~\ref{fig1-2} illustrates the locations of the event and Cauchy horizons for the MCRBH with varying $q$ parameters.

\begin{figure}[htb]
	\centering 
	\includegraphics[width=.325\textwidth]{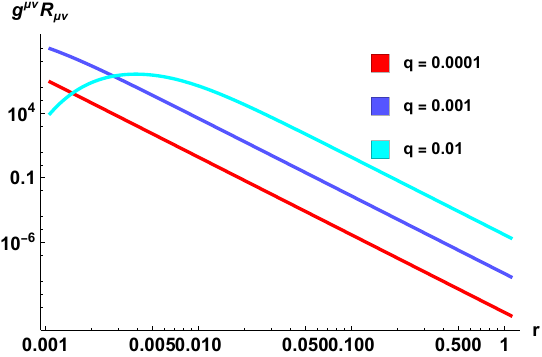}  
	\hfill
	\includegraphics[width=.325\textwidth]{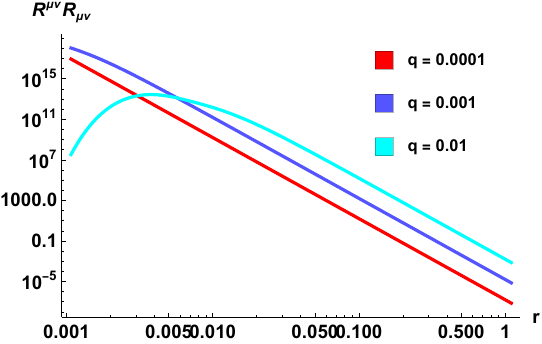}
	\hfill
	\includegraphics[width=.325\textwidth]{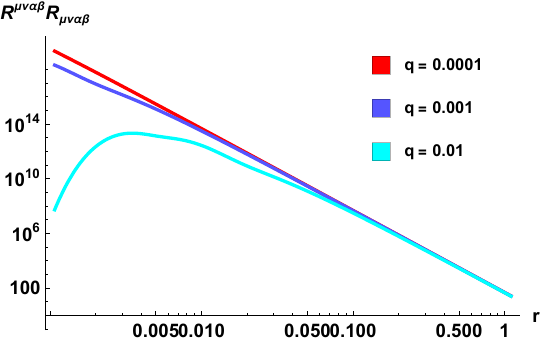}
	\caption{\label{fig1-3} 
Graphical representation of scalar invariants in MCRBH spacetime as a function of $r$ for varying values of the BH magnetic charge parameter $q$. Panels from left to right correspond to the Ricci scalar, to the Squared Ricci tensor, and to the Kretschmann scalar, respectively	
}
\end{figure}

The curvature invariants are mathematical quantities that provide insight into the characteristics of spacetime for a geometric structure such as a BH. Prominent scalar invariants comprise the Ricci scalar, the squared Ricci tensor, and the Kretschmann scalar (the square of the Riemann curvature tensor). Here, we first determine and examine these quantities
and then we will analyze their trends using Figure~\ref{fig1-3}.

The Ricci scalar for the BH metric \eqref{metric1}, considering the lapse function from Eq. \eqref{MetricFuncA}, can be determined as follows
\begin{equation}\label{RicciScalar}
R=g^{\mu\nu}R_{\mu\nu}=\frac{4M q^{2}}{r^{5}}\sech^{2}\left(\frac{q}{r}\right)\tanh \left(\frac{q}{r}\right)
\end{equation}
It is evident that when $q = 0$, the Ricci scalar likewise becomes zero.  
Next, we will visually examine the dependence on $q$ of various magnitudes.
In the left panel of Figure~\ref{fig1-3}, we reveal the Ricci scalar from Eq. \eqref{RicciScalar}, which is associated with the spacetime. In this way, we observe that the scalar invariant $R$ remains well-defined in the vicinity of the central point $r = 0$ when the parameter $q$ exceeds approximately $10^{-2}$. Therefore, we find that the BH solution remains regular at the center $(r\rightarrow0)$ for $q$ values approaching $10^{-2}$. Nonetheless, the Ricci scalar reaches infinity at $r\rightarrow0$ for $q < 10^{-2}$. This suggests that the spacetime surrounding the MCRBH does not possess a Ricci flat geometry.  Thus, it is evident that the Ricci scalar shows a rise as $q$ decreases.

Now, let us explore the squared Ricci tensor $\mathcal{R}$, whose expression is
\begin{equation}\label{squareRicciTensor}
\scalemath{0.95}{
\mathcal{R}=R^{\mu\nu}R_{\mu\nu}=\frac{4M^{2} q^{2}}{r^{10}}\sech^{6}\left(\frac{q}{r}\right)\left[-q^{2}+2r^{2}+\left(q^{2}+2r^{2}\right)\cosh \left(\frac{2q}{r}\right)-2qr\sinh \left(\frac{2q}{r}\right)\right]}.
\end{equation}
If $q = 0$, the Squared Ricci tensor is equal to zero. We present the analysis of Eq. \eqref{squareRicciTensor} using a graphical form. The quantity is depicted in the middle panel of  Figure~\ref{fig1-3}, revealing its well-defined behavior at the central point $r=0$ for $q \sim10^{-2}$. It is worth mentioning that the features of the square of the Ricci tensor are identical to those of the Ricci scalar.

The Kretschmann scalar, another scalar invariant, offers additional insights into spacetime curvature, notably distinguishing itself by not vanishing in a Ricci flat spacetime \cite{RayimbaevPoDU2022}. The formulation for the Kretschmann scalar \cite{CapozzielloJCAP2023,Capozziello2023} in the context of an MCRBH spacetime is as follows
\begin{equation}\label{KretschmannScalar}
\begin{split}
K=R^{\mu\nu\alpha\beta}R_{\mu\nu\alpha\beta}=\,&\frac{16M^{2}}{r^{10}}\left[-q^{4}\sech^{6}\left(\frac{q}{r}\right)+6r^{4}\left[1-\tanh\left(\frac{q}{r}\right)\right]\right.\\
&\left.+q^{2}\sech^{4}\left(\frac{q}{r}\right)\left[q^{2}+7r^{2}-4qr\tanh\left(\frac{q}{r}\right)\right]\right.\\
&\left.+r^{2} \sech^{2}\left(\frac{q}{r}\right)\left[-2q^{2}-6qr-3r^{2}+2q\left(q+3r\right)\tanh\left(\frac{q}{r}\right)\right]
\right].
\end{split}
\end{equation}
Observing the behavior of the Kretschmann scalar, in Figure~\ref{fig1-3} it is evident that for $q = 0$, it equals $\frac{48M^{2}}{r^{6}}$, 
and coincides with the Kretschmann scalar for the Schwarzschild case. Similar remarks apply to the properties of the Kretschmann scalar, mirroring the characteristics of other discussed scalar invariants.

\section{Geodesics around magnetically charged regular black hole}\label{Sec3}
\subsection{Null geodesic: Black hole shadow and constraints}
In this section, we initially explore the shadow radius of the Schwarzschild solution, which is modified with the inclusion of the additional parameter $q$ arising from the coupling of NLED and Einstein gravity in the weak-field limit. The boundary of the BH shadow, as seen by a distant observer, delineates the observable picture of the photon region by distinguishing between capture orbits and scattering orbits. The photon area is, practically, the edge of the spacetime region that, in the case of spherically symmetry spacetime, corresponds to the photon sphere. 
Next, we use the empirical data for M87* and  Sgr A* from the EHT collaboration , which are reported in Table \ref{tab3-1}, 
to put constraints on the $q$ parameter. 
To do so,  a starting point would be the Lagrangian $\mathcal{L}(x,\dot{x})=\frac{1}{2}g_{\mu \nu }\dot{x%
}^{\mu }\dot{x}^{\nu }$ for the geodesics of a spherically symmetric static spacetime metric, where:
\begin{equation}\label{Lagrangian1}
\mathcal{L}(x,\dot{x}) =\frac{1}{2}\left( -\text{A}(r)\dot{t}^{2}+\text{A}(r)^{-1}\dot{r}^{2} +r^{2}\left(\dot{\theta}^{2}+\sin^{2}\theta\dot{\varphi}^{2}\right)\right),
\end{equation}
so that $\text{A}(r)$ is given in Eq. \eqref{MetricFuncA} and the dot on top denotes
$d/d\lambda$, where $\lambda$ is the affine parameter.
For the sake of simplicity, let us focus on the movement of photons that are restricted to the equatorial plane of a BH. This means that the polar angle is set at a fixed value of $\theta=\frac{\pi}{2}$.
Since two conserved quantities, 
\begin{equation}\label{ConservedQunatityEL}
E=\text{A}(r)\dot{t}\quad \text{and} \quad L=r^{2}\dot{\varphi}, 
\end{equation}
which stand for energy and angular momentum, respectively, exist due to the coefficient of the metric equation cannot be found explicitly using the $t$ and $\varphi$ coordinates \cite{PerlickPR2022,FengEPJC2020,CapozzielloJCAP2023,Capozziello2023,UniyalPoDU2023, LambiaseEPJC2023, KhodadiPRD2022}.
For our purposes, it is more convenient to apply a first integral of the geodesic equation, that is, $2\mathcal{L}=0$ attributed to light. Therefore
\begin{equation}\label{GeoEq2}
-\text{A}(r)\dot{t}^{2}+\text{A}(r)^{-1}\dot{r}^{2} +r^{2}\dot{\varphi}^{2} = 0.
\end{equation}
By substituting the conserved quantities $E$ and $L$, of Eq. \eqref{ConservedQunatityEL} into Eq. \eqref{GeoEq2}, we obtain the orbit equation for photons as
\begin{equation}\label{OrbitEq1}
\left(\frac{dr}{d\varphi}\right)^{2} + V_{\text{eff}}= 0,
\end{equation}
where the effective potential provided by
\begin{equation}\label{veff}
V_{\text{eff}} = -r^{4} \left(\frac{1}{b^{2}}-\frac{\text{A}(r)}{r^2}\right), \qquad b=\frac{L}{E}.
\end{equation}
As the orbit equation only depends on the impact parameter $b$ at the trajectory's turning point, $r=r_{\text{Ph}}$, we have to determine the photon sphere radius, $r_{\text{Ph}}$, for the specified metric \eqref{metric1}. The two requirements $\frac{dr}{d\varphi} = 0$ and $\frac{d^{2}r}{d\varphi^{2}} = 0$ should be satisfied simultaneously along a circular photon orbit. By differentiating Eq. \eqref{OrbitEq1}, one can obtain the condition $\frac{d^{2}r}{d\varphi^{2}} = 0$. By solving the two equations simultaneously, we determine the equation representing the photon sphere radius in the following form
\begin{equation}\label{rph1}
	\left.\frac{d}{dr}\left(\frac{r^{2}}{\text{A}(r)}\right)\right|_{r=r_{\text{Ph}}} =0.
\end{equation}
Solving numerically Eq. \eqref{rph1}, we explore the impact of the deviation parameter $q$, corresponding to the BH magnetic charge, on the behavior of the photon sphere radius $r_{\text{Ph}}$. 
This is illustrated in Figure~\ref{fig1-2} by a numerical plot which reveals the location of the photon sphere radius in terms of $q$ values. 
This is significant because the critical impact parameter $b_{\text{C}}$ can be determined by $r_{\text{Ph}}$, the shadow cast and the behavior of the shadow radius depend on the photon sphere radius. 
Figure~\ref{fig1-2} shows that as the parameter $q$ grows, photon sphere radius and the critical impact parameter significantly decreases.
For the conventional Schwarzschild metric, it can be presented that these radii are $r_{\text{Ph}} = 3M$ and $b_{\text{C}} = 3\sqrt{3}M$, respectively.

In order to form the shadow, all light rays that travel into the past from the static observer's location at $(t_{\text{O}},r_{\text{O}},\theta_{\text{O}} = \frac{\pi}{2},\varphi_{\text{O}}=0)$ are taken into consideration. The angle at which they depart from the radial line is $\theta_{\text{S}}$ \cite{PerlickPRD2018}, satisfying 
 \begin{equation}
 \tan \theta_{\text{S}} =\left. \lim_{\Delta x \to 0} \frac{\Delta y}{\Delta x} = \sqrt{\frac{r^{2}}{B(r)}}\,\frac{d\varphi}{dr}\right|_{r=r_{\text{O}}}.
 \end{equation}
 This expression can also be represented as 
 \begin{equation}
 \sin \theta_{\text{S}} = \frac{b_{\text{C}}}{r_{\text{O}}}\sqrt{\text{A}(r_{\text{O}})}.
 \end{equation}
Therefore, for a static observer $r_{\text{O}}$, the BH shadow radius is
 \begin{equation}\label{shadowradius}
R_{\text{S}} =  b_{\text{C}}\sqrt{\text{A}(r_{\text{O}})}, \quad \text{or} \quad R_{\text{S}} = r_{\text{Ph}} \sqrt{\frac{\text{A}(r_{\text{O}})}{\text{A}(r_{\text{Ph}})}}.
\end{equation}
The shadow radius is obviously depends on the position of the observer, as demonstrated by equation \eqref{shadowradius}. We note that for a remote observer from a BH with an asymptotically flat metric, it reduces to $R_{\text{S}}= b_{\text{C}}$. The reason for this simplification is that $A(r_{\text{O}}) \approx 1$ at a significant distance from the BH
\cite{VagnozziCQG2023}.
This condition is met for the Schwarzschild metric if $M \ll r_{\text{O}} $, which means that the distance between the  observer and the BH is much greater than its gravitational radius. When comparing the distances of M87* and Sgr A* from us, which are $\approx 16.8 \,\text{Mpc}$ and $\approx 8 \text{kpc}$, respectively, to their gravitational radii, $r_{\text{g}} = \,\mathcal{O}(10^{-7})$ pc, we can see that this requirement is satisfied in both cases.

\begin{figure}[htb]
	\centering 
	\includegraphics[width=.45\textwidth]{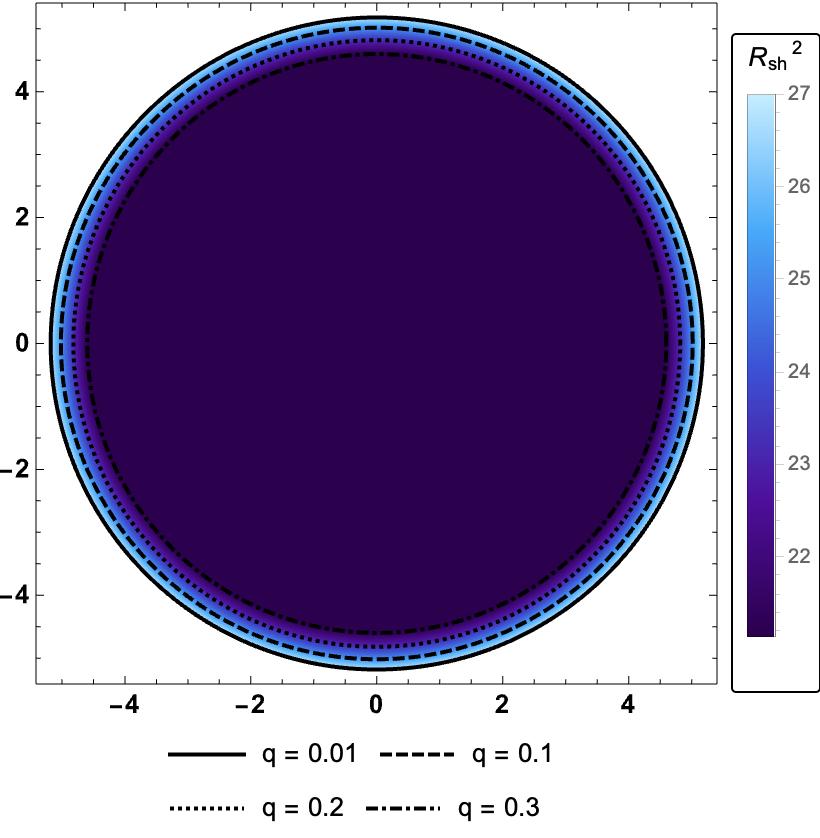}  
	\caption{\label{fig3-1} Observation of the MCRBH solution shadows from the perspective of an observer located at spatial infinity,  showcasing variations with different $q$ values.}
\end{figure}

We are now in an appropriate position to disclose the influence of the deviation parameter $q$ on the shadow radius size with spherical symmetry corresponding to the MCRBH solutions as observed by a viewer at spatial infinity. We reveal the shadow arising from a few optional values of $q$ in Figure~\ref{fig3-1}, which is linked to the lapse function \eqref{MetricFuncA2}. It is readily apparent that the inclusion of the term in the MCRBH metric, responsible for the non-linear electromagnetic field, significantly affects the variation in shadow size, causing it to shrink as $q$ increases.  Furthermore, some of the $q$ values chosen in Figure~\ref{fig3-1} lie within the following uncertainty bounds, and the associated allowable ranges will be estimated in light of the M87* and Sgr A* data, as shown below.

An interesting opportunity to do a precise test of gravitational theory in the strong and relativistic field regimes has been created by the BH shadow observations made by the EHT Collaboration. 
In addition, employing the Schwarzschild deviation parameter $\delta$ could be advantageous in determining restrictions on the parameters of a specific BH case \footnote{For details on determining the $1\sigma$ and $2\sigma$ confidence levels, interested readers are referred to Refs. \cite{KocherlakotaPRD2021,VagnozziCQG2023} and the references therein.}.
At this stage, we want to constrain the model parameter $q$ using the uncertainty provided by Refs. \cite{KocherlakotaPRD2021,VagnozziCQG2023}, and the shadow image observation data supplied by the EHT for M87* and Sgr A* \cite{AkiyamaL12019,AkiyamaL52019,AkiyamaL62019,AkiyamaL122022,AkiyamaL172022,DoScience2019,GravityCollaborationAA2022}.

\begin{figure}[htb]
	\centering 
	\includegraphics[width=.49\textwidth]{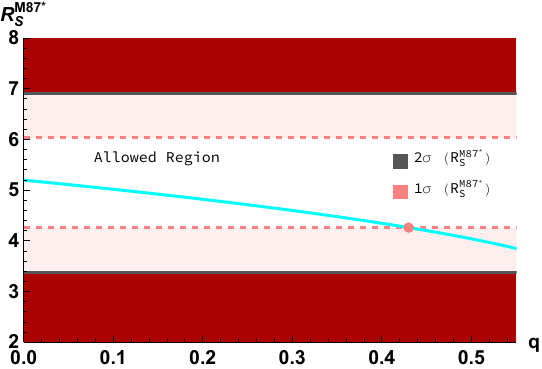}  
	\hfill
	\includegraphics[width=.49\textwidth]{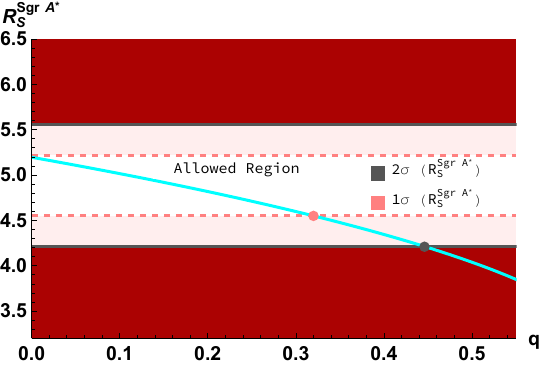} 
	\caption{\label{fig3-2} 
	Shadow radius of the MCRBH, characterized by the metric function in Eq. \eqref{MetricFuncA2} and expressed in units of the BH mass $M$, plotted versus the parameter $q$. The red shaded regions indicate values of $q$ inconsistent with stellar dynamics observations for M87* (left panel) and Sgr A* (right panel). The white and light pink shaded areas align with the EHT horizon-scale images of M87* and Sgr A* at $1\sigma$ and $2\sigma$ confidence levels, respectively. In the right panel, these shaded regions correspond to values of $q$ consistent with the averaged Keck and VLTI mass-to-distance ratio priors for Sgr A*.
	}
\end{figure}

Ref. \cite{AkiyamaL12019} reports that, for the M87*, the mass is $M_{\text{M87*}} = (6.5 \pm 0.7) \times 10^{9} \,\text{M}_{\odot}$, the angular diameter of the shadow is $\theta_{\text{M87*}}= 42 \pm 3 \,\mu\text{as}$, and the distance of the M87* from the Earth is $D_{\text{M87*}} = 16.8\pm0.8\,\text{Mpc}$. 
Considering Schwarzschild shadow deviations $\delta_{\text{M87*}} = -0.01\pm 0.17$ for M87*, where $\frac{R_{\text{S}}}{M} =3\sqrt{3}(1+\delta_{\text{M87*}})$ characterizes the shadow radius level, the shadow size of M87* is constrained to the range 
\begin{equation}\label{M87ShadowR}
4.26 \le \frac{R_{\text{S}}}{M} \le 6.03,
\end{equation} 
within the $1\sigma$ confidence region. 
In the recent EHT paper \cite{AkiyamaL122022}, data for Sgr A* indicates a shadow angular diameter of $\theta_{\text{Sgr A*}}= 48.7 \pm 7 \mu\text{as}$. The measurements infer the distance of Sgr A* from Earth as $D_{\text{Sgr A*}} = 8277\pm 9 \pm 33\, \text{pc}\, \text{(VLTI)},\,$$ 7953\pm50\pm32\, \text{pc}\,\text{(Keck)},\, 8150\pm 150\,\text{pc}\, \text{(VLBI)} $, with a BH mass of $M_{\text{Sgr A*}} = (4.297\pm 0.012 \pm 0.040)\times10^{6} \text{M}_\odot \, \text{(VLTI)},\,$$ (3.951\pm 0.047) \times 10^{6}\text{M}_\odot \, \text{(Keck)}, \, (4.0^{1.1}_{-0.6})\times10^{6}\text{M}_\odot \, \text{(EHT)}$. Therefore, using the Keck and VLTI measurements, the fractional deviation from the Schwarzschild expectation for Sgr A* are determined as $\delta_{\text{Sgr A*}} = -0.08^{+0.09}_{-0.09}$ $\,\text{(VLTI)}$, and $\delta_{\text{Sgr A*}} = -0.04^{+0.09}_{-0.10}\, \text{(Keck)}$ \cite{AkiyamaL172022,DoScience2019,GravityCollaborationAA2022}.
Considering the average of Keck and VLTI-based estimates, denoted as $\delta_{\text{Sgr A*}} \simeq 0.060^{+0.065}_{-0.065}\,\text{(Avg)}$, and employing $\frac{R_{\text{S}}}{M} =3\sqrt{3}(1+\delta_{\text{Sgr A*}})$ to characterize the shadow radius level, the shadow size of Sgr A* is confined to the range 
\begin{equation}\label{SgrAShadowR}
4.55 \le \frac{R_{\text{S}}}{M} \le 5.22,
\end{equation}  
within the $1\sigma$ confidence level. Figure~\ref{fig3-2} depicts the change of the shadow radius as a function of the parameter $q$ for M87* and Sgr A*, with uncertainties at $1\sigma$ and $2\sigma$ levels. Due to imposing the bound on the shadow radius, the numerical values for the lower bounds in $q$ are determined in Table \ref{tab3-1}. As expected, as $q$ increases, the shadow radius of MCRBH decreases.

\begin{table}[ht]
		\caption{Permissible values for the MCRBH parameter $q$, derived from the curves in Figure~\ref{fig3-2}, corresponding to the BH shadow radius consistent with the EHT horizon-scale images of M87* and Sgr A* at $1\sigma$ and $2\sigma$ confidence intervals.}
	\label{tab3-1}\centering
	\begin{tabular}{cccccc}
		\toprule
		{$q$} &\multicolumn{2}{c}{$1\sigma$}& & \multicolumn{2}{c}{$2\sigma$} \\
		\cmidrule{2-3} \cmidrule{5-6}
		
		{} & {Upper} & {Lower} &{}& {Upper} & {Lower} \\
		\midrule
		$\text{M87*}$ & -- & $0.43$ & & -- & -- \\
		$\text{Sgr A*}$ & -- & $0.32$ & & -- & $0.45$ \\
		\bottomrule
	\end{tabular}
\end{table}

From Figure~\ref{fig3-2} and Table \ref{tab3-1}, it is evident that the EHT-derived shadow radius for the BHs M87* and Sgr A* imposes constraints on the reduction of their corresponding BH shadow sizes. Values exceeding $q \simeq 0.43$ and $q \simeq 0.32$ are excluded within the $1\sigma$ confidence level for M87* and Sgr A*, respectively. Similarly, at the $2\sigma$ confidence level, this restriction on the shadow size reduction of the MCRBH rules out values beyond $q \simeq 0.45$ exclusively in the case of Sgr A*.
Therefore, we can conclude that the data for Sgr A* provides more robust constraints on the BH parameter $q$. This is evident from a specific point that intersects the lower bound within the $2\sigma$ uncertainty range.
We observe that at the vicinity of the BH, the BH parameter $q$ is effective, as revealed by its notable impact on the photon sphere radius. Therefore, the shadow cast can be influenced, even when observed from a distance. 
Thereby, based on the tabulated constraints for $q$, we deduce that M87* and Sgr A*  BHs could potentially be MCRBHs, given the current precision of astrophysical data.

\subsection{Time-like geodesic: ISCO and MBO radii}
\begin{figure}[htb]
	\centering 
	\includegraphics[width=.32\textwidth]{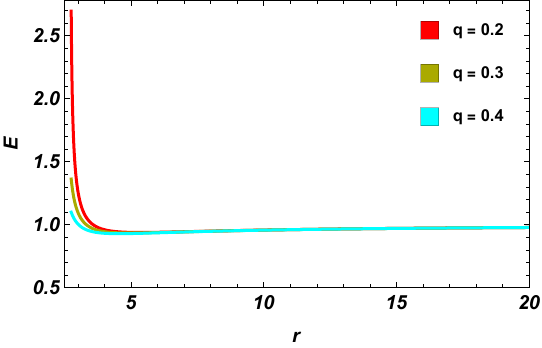}  
	\hfill
	\includegraphics[width=.32\textwidth]{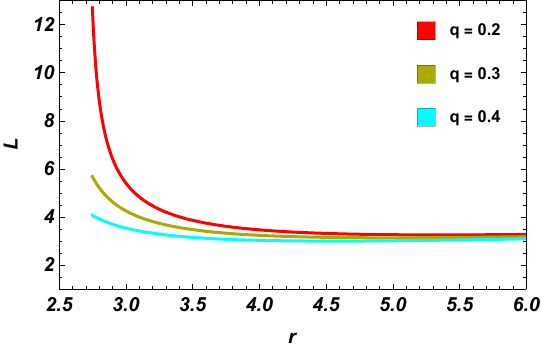}
	\hfill
	\includegraphics[width=.32\textwidth]{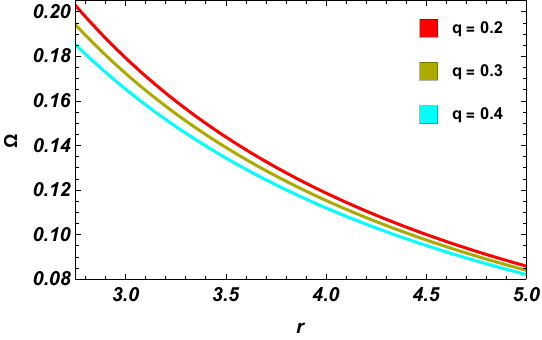}
	\caption{\label{fig3-3} 
	Variation of the specific energy (left), the angular momentum (middle) and the angular velocity (right) of orbiting particles with three different values of $q$.}
\end{figure}

\begin{figure}[htb]
	\centering 
	\includegraphics[width=.56\textwidth]{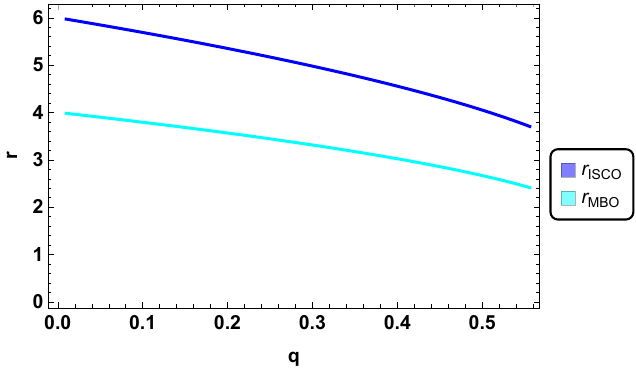}  
	\caption{\label{fig3-4} 
		The plot shows the behavior of the MBO radius $r_{\text{MBO}}$ and ISCO radius $r_{\text{ISCO}}$ with varying $q$. 
                   }
\end{figure}

In the framework of the coupled system involving Einstein gravity and NLED, the Schwarzschild spacetime, which traditionally exhibits the innermost stable circular orbit (ISCO) and marginally bound orbit (MBO) at $r_{\text{ISCO}} = 6M$ and $r_{\text{MBO}} = 4M$ for massive particles, respectively, undergoes variations. 
Therefore, by considering time-like orbits in this scenario, we are now going to focus on exploring the behavior of ISCO and MBO radii for an MCRBH as a function of the deviation parameter $q$ (see Figure~\ref{fig3-4}).
Thus, Eq. \eqref{GeoEq2} can be rearranged to read 
\begin{equation}\label{GeoEq22}
-\text{A}(r)\dot{t}^{2}+\text{A}(r)^{-1}\dot{r}^{2} +r^{2}\dot{\varphi}^{2} = -1.
\end{equation}
 by considering that $2\mathcal{L}=-1$ for a massive particle. By applying the conserved quantities $E$ and $L$ given in \eqref{ConservedQunatityEL}, and narrowing our focus to equatorial orbits, the effective potential associated with the corresponding orbit equation becomes
\begin{equation}\label{veffTL}
	V_{\text{eff}} = -r^{4} \left(\frac{E^{2}}{L^{2}}-\frac{\text{A}(r)}{L^2}-\frac{\text{A}(r)}{r^{2}}\right).
\end{equation}
Making use of the prerequisites $V_{\text{eff}} = 0$ and $V_{\text{eff,r}} = 0$ to achieve stable circular orbits, the specific energy $E$, specific angular momentum $L$, and angular velocity $\Omega$ of particles in motion within the equatorial plane while the BH gravitational potential is present, can be obtained as follows
\begin{subequations}
\begin{align}
E &= \frac{\sqrt{2}\, \text{A}(r)}{\sqrt{2 \text{A}(r)-r \text{A}'(r)}}, \label{E}\\
L &= \frac{r^{3/2}\, \sqrt{\text{A}'(r)}}{\sqrt{2 \text{A}(r)-r \text{A}'(r)}},\label{L}\\
\Omega &= \frac{d\varphi}{dt} = \sqrt{\frac{\text{A}'(r)}{2r}}.
\end{align}
\end{subequations}
It is evident from Eqs. \eqref{E} and \eqref{L} that the specific energy and angular momentum must both be real provided that the condition $2 \text{A}(r)-r \text{A}'(r) >0$ is satisfied.
For various values of the BH parameter $q$, the plots for the specific energy, specific angular momentum and angular velocity are displayed in Figure~\ref{fig3-3}. All quantities decrease as $q$ increases; this is determined by the variation in the effective potential with respect to $q$.

The stability of the bounded orbits is denoted by the values of $V_{{\rm eff},rr}$. Bounded orbits may be stable or unstable. One possible interpretation is that a bounded orbit is unstable if $V_{{\rm eff},rr}< 0$, indicating that it corresponds to a maximum point in the effective potential, where even a minor perturbation would cause it to destabilize. Alternatively, if $V_{{\rm eff},rr}> 0$, then a little perturbation would cause tiny oscillations around the orbit; indicating a stable orbit. 
We concentrate on two pivotal bounded orbits: ISCO and MBO. ISCO, the minimal stable orbit encircling a compact object, aligns with the turning point of the effective potential when $V_{\text{eff,rr}} = 0,$ providing insights into BH accretion disks. MBO, representing the critical bound orbit with energy $E = 1,$ demarcates the boundary between bounded $E < 1$ and unbounded $E > 1$ orbits, crucial for elucidating the dynamics of star clusters surrounding supermassive BHs \cite{WillCQG2012}.
Now, to determine the radius of the ISCO, $r_{\text{ISCO}}$, of the MCRBH, we must satisfy the following 
conditions must be satisfied: $V_{{\rm eff}}=0=V_{{\rm eff},r}$; and $V_{{\rm eff},rr}=0$. By combining them, the solution to the equation 
\begin{equation}\label{ISCOradius}
\frac{2r \text{A}'(r)^{2}}{3 \text{A}'(r)-r \text{A}''(r)}-\text{A}(r) = 0,
\end{equation}
is obtained as $r_{\text{ISCO}}$.
Moreover, by setting $E=1$ in Eq. \eqref{E}, we arrive to the expression 
\begin{equation}\label{MBOradius}
r \text{A}'(r)+2\text{A}(r)(\text{A}(r)-1)=0,
\end{equation} 
allowing us to determine the position of the MBO. By numerically solving Eqs. \eqref{ISCOradius} and \eqref{MBOradius}, we obtain the radii $r_{\text{MBO}}$ and $r_{\text{ISCO}}$ for various BH parameters $q$, as depicted in Figure~\ref{fig3-4} through a numerical plot. 
The plot reveal that with increasing $q$, both the radii of the MBO and ISCO decrease.

\section{Thin accretion disks around MCRBH}\label{Sec4}
Within the framework of NLED in the weak-field limit coupled with GR, we will investigate the impact of the MCRBH parameter $q$ on the radiation emanating from the accretion disk {\color{cyan}\cite{ShakuraAA1973,Novikov1973,PageApJ1974,BambiApJ2011,BambiApJ2012,ChenPLB2011,PerezAA2013, CollodelApJ2021,BoshkayevPRD2021}} and the shadow cast by the BH employing the thin accretion disk model {\color{cyan}\cite{UniyalPoDU2023,BromleyApJ1997,YuanARAA2014,BambiPRD2013,JaroszynskiAA1997,GrallaPRD2019,ZengEPJC2020,PengCPC2021,ZengPRD2023,GuoPRD2022,WangPRD2023,WaliaJCAP2023,FathiEPJC2023,LiEPJC2021,LambiaseJCAP2023,ZengEPJC2020-2,HuEPJC2022,ShaikhMNRAS2019}}. To do so, we first explore the time-averaged energy flux $F$, the differential of the luminosity $\mathcal{L}_{\infty}$ and the disk temperature $T$. Additionally, we will determine the thickness of the rings by visualizing the light rings and shadows for three distinct disk emission profiles. In addition to optically thin disk accretion, which is regarded as just a background light source, we will examine the shadow and observed luminosity of the MCRBH surrounded by other accretion models, namely static and infalling spherical accretion flows.

\subsection{Characteristics of relativistic thin accretion disk}
\begin{figure}[htb]
	\centering 
	\includegraphics[width=.325\textwidth]{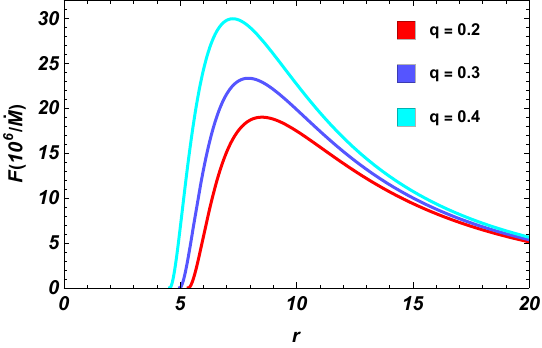}  
	\hfill
	\includegraphics[width=.325\textwidth]{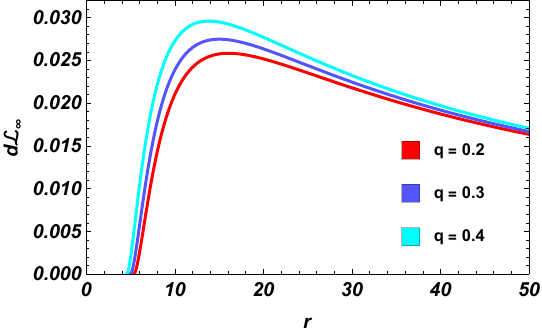}
	\hfill
	\includegraphics[width=.325\textwidth]{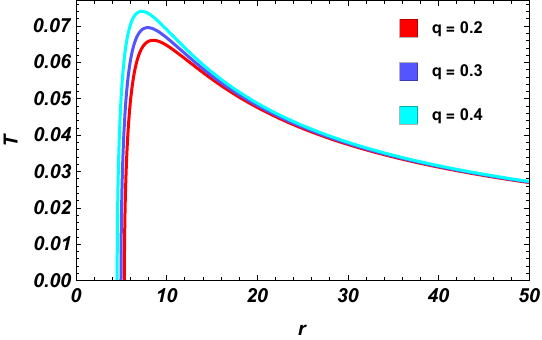}
	\caption{\label{fig4-1} 
The radiant energy flux per unit disk accretion rate (left panel), the differential luminosity at infinity (middle panel), and the temperature (right panel) of a thin accretion disk for different values of the parameter $q$.}
\end{figure}

The relativistic formulation of the thin accretion disk model, an extension of the renowned Shakura–Sunyaev model \cite{ShakuraAA1973}, was developed in the early 1970s by Novikov and Thorne \cite{Novikov1973} and Page and Thorne \cite{PageApJ1974}. The equations that regulate the phenomenon are derived from a set of simple, yet reasonably justified suppositions. 
They made assumptions regarding the background spacetime geometry, considering it to be stationary, axially symmetric, asymptotically flat, and symmetrically reflective about the equatorial plane. Furthermore, they postulated that the central plane of the disk aligns with the equatorial plane of the BH.
They also proposed that the disk is thin, meaning that the height of the accreting disk is insignificant in comparison to its horizontal extension at any given radius, $H\ll r$, where $H$ denotes the disk's maximum half-thickness, and that its central plane is exactly in the equatorial plane.
The self-gravity of the disk is also assumed to be negligible, meaning that the mass of the disk does not affect the background metric. 
It is important to note that the entire disk is assumed to be in a state of local hydrodynamical equilibrium at each point. The pressure gradient and vertical entropy gradient within the disk are thought to be insignificant. The cooling in the disk is assumed to be effective enough to prevent the accumulation of heat created by stresses and dynamic friction within the disk. So, this cooling helps to stabilize the disk's vertically thin structure. 
In this way, the mass accretion rate $(\dot{M})$ is taken to be constant over time and independent of the radial coordinate since the disk is thought to be in a stable state. The inner boundary of the disk coincides with the ISCO, while the material distant from the BH is assumed to exhibit Keplerian motion.
In this steady-state accretion disk scenario, the accreting matter within the disk can be characterized by the energy-momentum tensor of an anisotropic fluid as follows:
\begin{equation}\label{EnergMomTen}
T^{\mu\nu} = \varepsilon_{0}u^{\mu}u^{\nu} +u^{(\mu}q^{\nu)}+t^{\mu\nu},
\end{equation}
where, defined in the averaged rest-frame of the orbiting particle with four-velocity $u^{\mu}$, each of $\varepsilon_{0}$, $q^{\mu}$, and $t^{\mu\nu}$ stand for respectively, the rest mass density, the energy flow vector, and the stress tensor of the accreting matter. Within this frame, it holds that $u_{\mu}q^{\mu} = 0 = u_{\mu}t^{\mu\nu}$, as both $q^{\mu}$ and $t^{\mu\nu}$ are orthogonal to $u^{\mu}$. 
By applying the conservation laws to the rest mass (viz.,~$\nabla_{\mu}(\varepsilon_{0}u^{\mu}) = 0$), energy $E$ (viz.,~$\nabla_{\mu} E^{\mu} = 0$), and angular momentum $L$ (viz.,~$\nabla_{\mu} J^{\mu} = 0$), three time-averaged radial structure equations of the thin disk surrounding the MCRBH can be derived as follows:
\begin{subequations}
\begin{align}
&\,\,\dot{M} = -2\pi \sqrt{-G}\, \Sigma(r) u^{r} = \text{Const},\\
&\left[\dot{M} E -2\pi \sqrt{-G} \Omega W^{r}_{\phi}\right]_{,r} = 2\pi \sqrt{-G} F(r) E,\label{CLE}\\
&\left[\dot{M} L -2\pi \sqrt{-G} \Omega W^{r}_{\phi}\right]_{,r} = 2\pi \sqrt{-G} F(r) L.\label{CLM}
\end{align}
\end{subequations}
Here, the time derivative, denoted by a dot, is taken with respect to the time coordinate $t$. For the case considered here, $G=g_{tt}g_{rr}g_{\varphi\varphi}$. The averaged rest mass density $\Sigma(r)$ and the averaged torque $W_{\varphi}^{r}$ are expressed as follows:
\begin{equation}\label{WS}
\Sigma(r) = \int_{-H}^{H} \langle\varepsilon_{0}\rangle\, dz, \qquad W_{\varphi}^{r} =  \int_{-H}^{H} \langle t_{\varphi}^{r}\rangle\, dz,
\end{equation}
where the symbol $\langle t_{\varphi}^{t}\rangle$ represents the value of the stress tensor's $(\varphi,r)$ component, averaged over a characteristic time interval $\Delta t$ and an azimuthal angle of $\Delta\varphi = 2\pi$. 
By employing the energy-angular momentum relation for geodesic orbits, which states that $E_{,r} = \Omega L_{,r}$, it is possible to exclude $W_{\varphi}^{r}$ from Eq. \eqref{CLM} and \eqref{CLE}. This allows us to derive the expression for the time-averaged energy flux $F(r)$ emitted from the surface of an accretion disk surrounding the compact object, which is provided by
\begin{equation}\label{EnergyFlux}
F(r) = - \frac{ \dot{M}_{0}\Omega_{,r}}{4\pi \sqrt{-G}(E-\Omega L)^{2}}\int^{r}_{r_{\text{ISCO}}}(E-\Omega L)L_{,r}\,dr.
\end{equation}
It is important to note that the energy flux $F$ is not directly observable, as it represents a local quantity measured in the rest frame of the disk. As such, a combination of energy and angular momentum conservation laws yields a more compelling observational quantity: the differential luminosity $\mathcal{L}_{\infty}$, which denotes the energy per unit time observed by an observer located at infinity. This quantity can be determined using the energy flux $F$ according to the following expression
\begin{equation}\label{Luminosity}
\frac{d\mathcal{L}_{\infty}}{d \text{ln}\,r} = 4 \pi r \sqrt{-G}\, E F(r).
\end{equation}
Both of these characteristics quantify the magnitude of radiation emitted by the corresponding disk at a specific radius $r$. 
Considering the thermal equilibrium of the disk, as mentioned earlier, the emitted radiation can be treated as black body radiation, with the temperature expressed by 
\begin{equation}\label{Temp}
T(r) = \sigma^{-\frac14}F^{\frac14},
\end{equation} 
where $\sigma$ represents the Stefan-Boltzmann constant.
Left, middle and right panels of Figure~\ref{fig4-1} display the plots for the energy flux per unit mass accretion rate, the differential luminosity per unit mass accretion rate, and the temperature of the thin accretion disk $\sigma^{1/4}T /\dot{M}^{1/4}$ as functions of the radial coordinate.

\subsection{Rings and black hole shadows in various accretion disk flow scenarios} 
The characteristics of a BH shadows and rings are influenced not only by its spacetime but also by the properties of the BH accretion disk. This section explores the observed shadow images, rings, and optical appearance of the MCRBH with both thin disk-shaped and spherical accretion models.

\subsubsection{Thin accretion disk flow}
In this subsection, we will investigate how incoming rays are categorized and how optically and geometrically thin accretion disks, resting on the equatorial plane surrounding the black hole and observable face-on from the north pole direction, illuminate images of MCRBH.
In this scenario, incoming rays may intersect with the accretion disk at various times, leading to different contributions to the overall observed intensity. Therefore, our approach involves initially categorizing these incoming rays, followed by revealing black hole images through the analysis of the total observed intensity across three standard accretion profiles.

\paragraph{Direct emission, lensed ring and photon ring:}

Here, we will explore how photon trajectories influence the optical appearance of the BH shadow and the observed emission profile, within the context of a MCRBH. We investigate this phenomenon by considering an optically thin and geometrically thin disk-shaped accretion flow as a case study. Our assumptions include isotropic emission in the rest frame of static worldlines, with the disk situated in the equatorial plane and the observer positioned at a significant distance from the BH in the north pole direction.
Based on Ref. \cite{GrallaPRD2019}, a key characteristic of a BH surrounded by a thin disk accretion flow is the presence of a lensed ring and photon ring encircling the BH shadow. Initially, we examine the trajectory of a light ray as it orbits the BH, which can be described by the variation of the radial coordinate with the azimuthal angle $\varphi$.
By taking $u=\frac{1}{r}$, the orbit equation can now be rearranged as
\begin{equation}\label{OrbitEq2}
	\left(\frac{du}{d\varphi}\right)^{2}=G(u)\equiv \frac{u^{4}\text{D}(u)^{2}}{\text{A}(u)\text{B}(u)}\left(\frac{1}{b^{2}}-\frac{\text{A}(u)}{\text{D}(u)}\right),
\end{equation}
in which, $\text{A}(u) = \text{B}(u)^{-1}$ and $\text{D}(u) = \frac{1}{u^{2}}$. 
To facilitate the use of the ray-tracing code for demonstrating the bending of light around the MCRBH, we employ the lapse function provided in Eq. \eqref{MetricFuncA2}. Given that Eq. \eqref{OrbitEq2} is a function of the impact parameter $b$, we can thus expect that the geometry of geodesics relies only on the roots of the equation $G(u)=0$.
In light of this, the following describes how light rays move around a BH: 
I) If $b>b_{\text{C}}$, the light ray will deflect at $u_{\text{Min}}$\footnote{The subscript $\text{Min}$ in $u_{\text{Min}}$, denoting the turning point, indicates the smallest positive real root of the equation $G(u)=0$.}, the radial position where $G(u)|_{u=u_{\text{Min}}} =0$, and go indefinitely away from the BH; 
II) When $b<b_{\text{C}}$, the light ray is always caught by the BH and is unable to go indefinitely; 
III) The photons are in a rotating state around the BH when $b=b_{\text{C}}$; neither falling into the BH nor escaping from it.
Thus, based on Eq. \eqref{OrbitEq2}, the total change in azimuthal angle $\varphi$ for a given trajectory with impact parameter $b$ can be determined as
\begin{equation}
\varphi = 
\begin{cases}
	2\int_{0}^{u_{\text{Min}}}\frac{du}{\sqrt{\frac{1}{b^{2}}-u^{2}\text{A}(u)}},\quad b>b_{\text{C}},\\
	\int_{0}^{u_{+}}\frac{du}{\sqrt{\frac{1}{b^{2}}-u^{2}\text{A}(u)}},\qquad\,\, b<b_{\text{C}}, 
\end{cases}
\end{equation}
so that for $b<b_{\text{C}}$, our focus lies on the trajectory beyond the horizon \cite{GrallaPRD2019}. Here, $u_{+}$ is associated with the radius of the event horizon as $u_{+}=\frac{1}{r_{+}}$.
The number of occurrences at which the light intersects the thin disk accretion gives the classification of the rings \cite{GrallaPRD2019}. As the total number of light orbits is defined as $n(b) = \frac{\varphi}{2\pi}$, the following describes the trajectories of light rays emitted from the north pole (far right of the trajectory plots):
I) Direct emission ($n< \frac{3}{4}$): the thin accretion disk is only intersected once by the light trajectories;
II) Lensed ring ($\frac{3}{4} <n< \frac{5}{4}$): the thin accretion disk is intersected twice by the light trajectories;
III) Photon ring ($n> \frac{5}{4}$): there are at least three intersections between the light trajectories and the thin accretion disk. 
In Table \ref{TabAccDisk1}, we illustrate the change in the BH shadow as $q$ increases, indicating the range of $b$ values for direct emission, lensed ring emission, and photon ring emission of the MCRBH for a couple of $q$ values, which correspond to the BH magnetic charge parameter.
One can observe that the lensed rings and photon rings become thicker as the parameter $q$ increases. 
\begin{table}[ht]
	\caption{Regions of direct rays, lensing rings, and photon rings for two
		different values of the MCRBH parameter $q$.}
	\label{TabAccDisk1}\centering
	\begin{tabular}{cccc}
		\toprule
		MCRBH parameter & $q=0.01$ & $q=0.1$ \\[0.5ex] \midrule
		Direct rays & $b<4.997$ & $b<4.824$ \\
		$\left(n < \frac{3}{4}\right)$ & $b>6.153$ & $b>6.014$ \\ [1.2ex]
		Lensing rings & \quad$4.997<b<5.170$ & \quad $4.824<b<5.006$ \\ 
		$\left(\frac{3}{4}<n<\frac{5}{4}\right)$ & \quad $5.211<b<6.153$ & \quad $%
		5.051<b<6.014$ \\ [1.2ex]
		Photon ring $\left(n>\frac{5}{4}\right)$ & \quad $5.170<b<5.211$ & \quad $%
		5.006<b<5.051$ \\[1ex] \bottomrule
	\end{tabular}%
\end{table}

\begin{figure}[htb]
	\centering 
	\includegraphics[width=.49\textwidth]{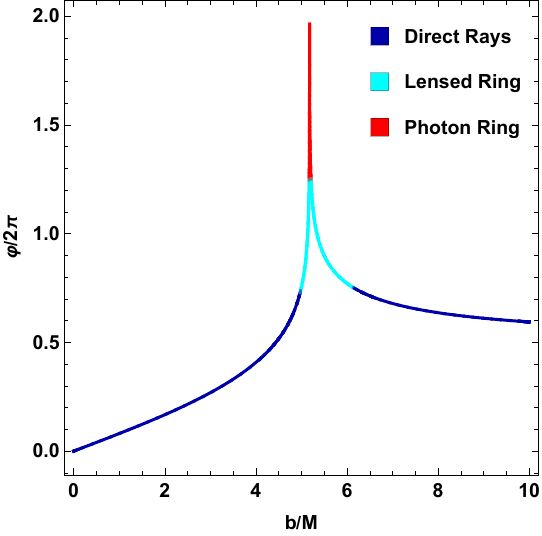}  
	\hfill
	\includegraphics[width=.49\textwidth]{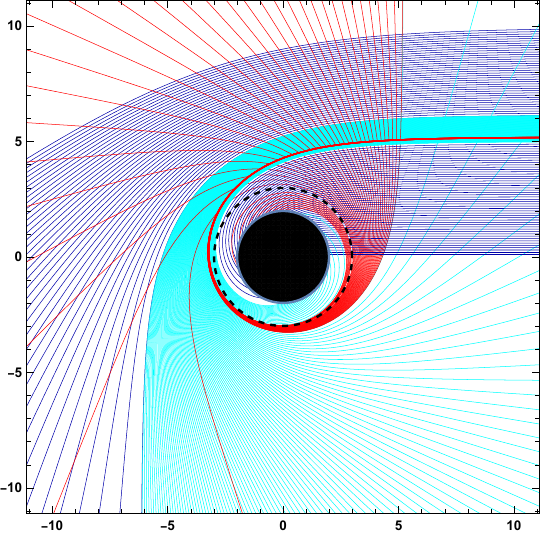}
	\\ [3ex]
	\includegraphics[width=.49\textwidth]{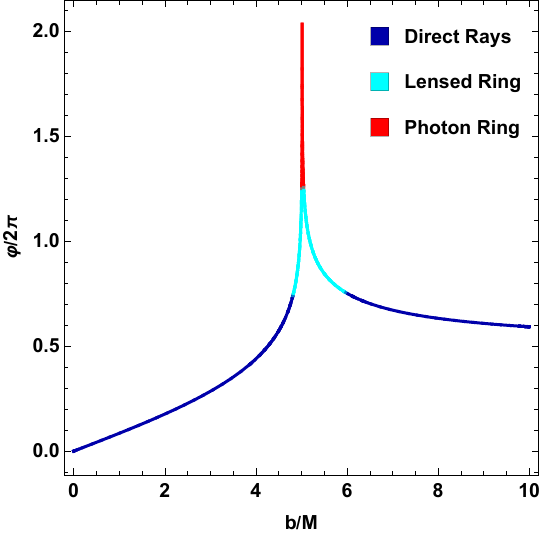}
	\hfill
	\includegraphics[width=.49\textwidth]{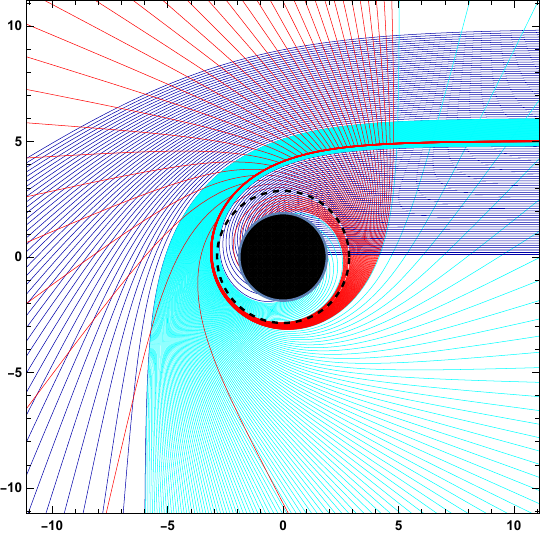}
		\caption{\label{fig4-2} 
		The left panels display the connection between the impact parameter $b$ and the total number of photon orbits $n$ for MCRBHs with varying values of $q$. The lines colored in dark blue, cyan, and red denote the direct emissions, lensed ring emissions, and photon ring emissions, respectively. 
		In the right panels, we observe the trajectories of photon rays corresponding to direct emissions (dark blue), lensed ring emissions (cyan), and photon ring emissions (red). The black dashed lines and black disks represent the photon ring and the event horizon, respectively. 
		The top row corresponds to $q = 0.01$, while the bottom row corresponds to $q = 0.1$.
	}
\end{figure}

In Figure~\ref{fig4-2}, the left panels display the total number of orbits as a function of the impact parameter $b_{\rm C}$ for $q = 0.01$ (top row) and $q =0.1$ (bottom row). We see that there is no significant disparity between the Schwarzschild BH and the MCRBH, indicating that the parameter linked to the BH magnetic charge has minimal influence on the classification of light trajectories \cite{GrallaPRD2019}. Additionally, when the impact parameter approaches the critical value $b\pm b_{\rm C}$, the photon orbit exhibits a narrow peak in the $(b,\varphi)$ plane. Subsequently, as $b$ increases, the photon trajectories consistently manifest as direct emissions across all cases.
From Table \ref{TabAccDisk1} and Figure~\ref{fig4-2}, it is evident that increasing the MCRBH parameter $q$ results in broader ranges of photon and lensed rings emissions, depicted by the red and cyan curves, respectively. Specifically, for $q$ on the order of $10^{-1}$, both photon ring and lensed ring emissions exhibit wider ranges of impact parameter compared to those of the Schwarzschild BH \cite{GrallaPRD2019}.
Increasing $q$ leads to a corresponding increase in the contribution to the brightness of both lensed and photon rings. Besides, the trajectories of light in polar coordinates are depicted in the right panels of Figure~\ref{fig4-2}, where the black disks represent the BHs, and the dashed black lines denote the photon ring.
In the following, we explore the observed emission intensity of the accretion disk within the context of NLED coupled with GR in the weak-field limit.

\paragraph{Transfer functions of MCRBH:}

\begin{figure}[hbt]
	\centering 
	\includegraphics[width=.49\textwidth]{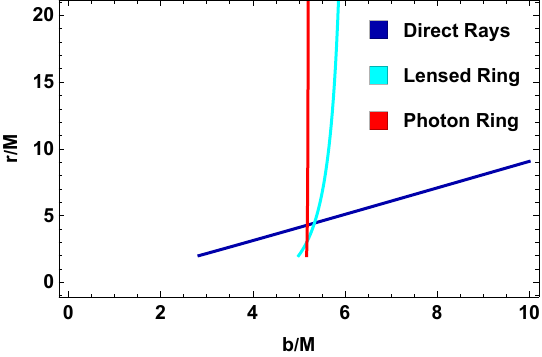}  
	\hfill
	\includegraphics[width=.49\textwidth]{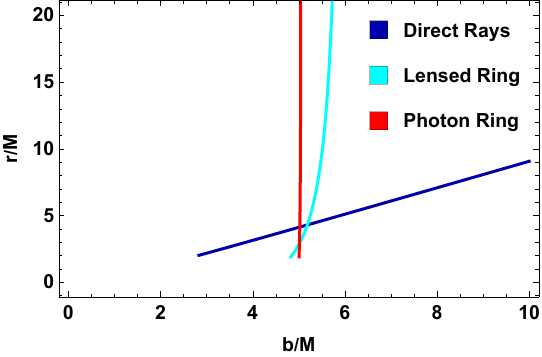}
	\caption{\label{fig4-3} 
		The plot of the first three transfer functions of MCRBHs for two different values of $q$ ( for the left panel $ q =0.01$ and for the right panel $ q =0.1$): the dark blue, cyan, and red lines represent respectively the first, second, and third transfer functions, corresponding to direct emissions, lensed ring emissions, and photon ring emissions.
	}
\end{figure}

Each time a light ray traverses the thin accretion disk, it extracts energy from it. Therefore, the total energy extracted depends on the number of passes through the thin disk.  
As a result, an observer at infinity will see a light ray whose intensity is proportional to the number of passes through it. 
We will then consider the relationship between the observed intensity of light and the emitted intensity, taking into account a distant static observer positioned at the north pole and a thin accretion disk situated at the equatorial plane of the BH. Furthermore, we will assume isotropic emission from the thin accretion disk for the static observer. Given this, the specific intensity and frequency of the emission are denoted as $I_{\rm Em}^{\nu} (r)$ and $\nu$ respectively, while the observed specific intensity and frequency are indicated as
$I_{\rm Obs}^{\nu'} (r)$ and $\nu' = \sqrt{\text{A}(r)}\,\nu$, where $\sqrt{A(r)}$ can be considered as a redshift factor and will be represented by $g_{\rm RF}$.
Ignoring absorption, according to Liouville's theorem, one can find that $I_{\rm Em}^{\nu} (r)/\nu^3$ is conserved along a ray \cite{BromleyApJ1997}. Thus, the specific intensity received by the observer with emission frequency 
$\nu$ is given by \cite{GrallaPRD2019,ZengEPJC2020,PengCPC2021,ZengPRD2023,GuoPRD2022,WangPRD2023}
\begin{equation}\label{Iobs&Imb}
I_{\rm Obs}^{\nu'} (r) = \text{A}^{3/2}(r) I_{\rm Em}^{\nu} (r).
\end{equation}
In turn, the overall observed intensity $I_{\rm Obs} (r)$ can be determined by integrating $I_{\rm Obs}^{\nu'} (r)$ across all observed frequencies, and can be expressed as 
\begin{equation}
I_{\rm Obs}(r) = \int I_{\rm Obs}^{\nu'} (r)\, d\nu' = \text{A}^{2}(r) I_{\rm Em} (r),
\end{equation} 
in which the total emitted intensity from the thin disk accretion flow, that is, $I_{\rm Em}$, is given by $I_{\rm Em} = \int I_{\rm Em}^{\nu}\, d\nu$.
As previously stated, energy is extracted by the light as it traverses the thin disk. As a result, the observer should get the intensity which is equal to the sum of the luminosities at each intersection point, that is, 
\begin{equation}\label{total_Iobs}
I_{\rm Obs}(r) = \sum_{m} \text{A}^{2}(r) I_{\rm Em} (r)|_{r=r_{\rm m}(b)},
\end{equation}
with $r_{m}(b)$ being the transfer function, 
as a transferring or mapping from the impact parameter $b$ of the light ray to the radial coordinate of the $m$th intersection between the light and the thin disk accretion flow. Furthermore, its slope $dr/db$ signifies the demagnification factor at each $b$.
In Figure~\ref{fig4-3}, we depict the first three transfer functions corresponding to $m = 1,2,3$ with respect to $b$ for two distinct values of $q = 0.01$ and $0.1$. These curves exhibit varying slopes, reflecting the demagnification factor and providing insight into the extent of demagnification observed in the image.
The dark blue curves depict the transfer function for direct emission, with an average slope close to one, suggesting it represents a direct image of the redshift source. The cyan curves represent the transfer function for lensed ring emission, exhibiting an average slope greater than one, indicating that it appears highly demagnified to the observer. The red curves correspond to the transfer function for photon ring emission, with a slope approaching infinity, implying that the observed photon ring is significantly demagnified.
The average slope of the transfer functions of the three types of light rays reveals their respective contributions to the total flux. The primary contribution to this total flux is from the direct emission rays, which have the smallest average slope. The lensed ring, with a significant average slope, also contributes. However, the photon ring, with an average slope approaching infinity, can be ignored due to its negligible contribution. Additionally, from Figure~\ref{fig4-3}, we observe that the transfer function moves to the left as $q$ grows. Furthermore, as $q$ grows, the average slope of the transfer functions for the lensed and photon rings also somewhat increases, suggesting that the contribution of these two kinds of light rays to the total flux increases.

\paragraph{The astronomical appearance of the MCRBH surrounded by thin disk accretion:}
Now we will focus on the observational characteristics of the MCRBH by considering three different inner radii at which the accretion flow ceases to emit radiation. One of the well-known relativistic effects is that the innermost stable circular orbit $r_{\rm ISCO}$ serves as the boundary between test particles orbiting the BH and those falling into it. We adopt the $r_{\rm ISCO}$ radius as the position where radiation emission stops.
The luminosity intensity of the shadow decreases exponentially as the accretion radiation ceases.

\begin{figure}[htb]
	\centering 
	\includegraphics[width=.32\textwidth]{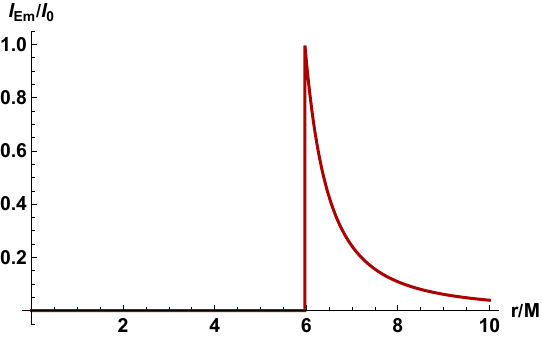}   
	\hfill
	\includegraphics[width=.32\textwidth]{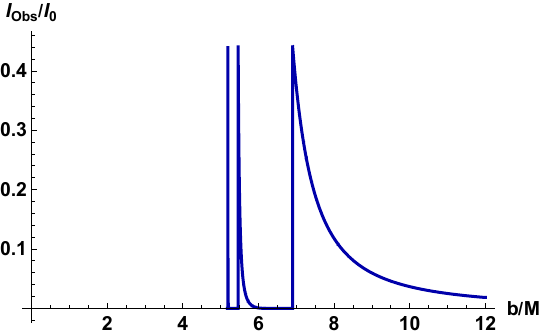}
	\hfill
	\includegraphics[width=.32\textwidth]{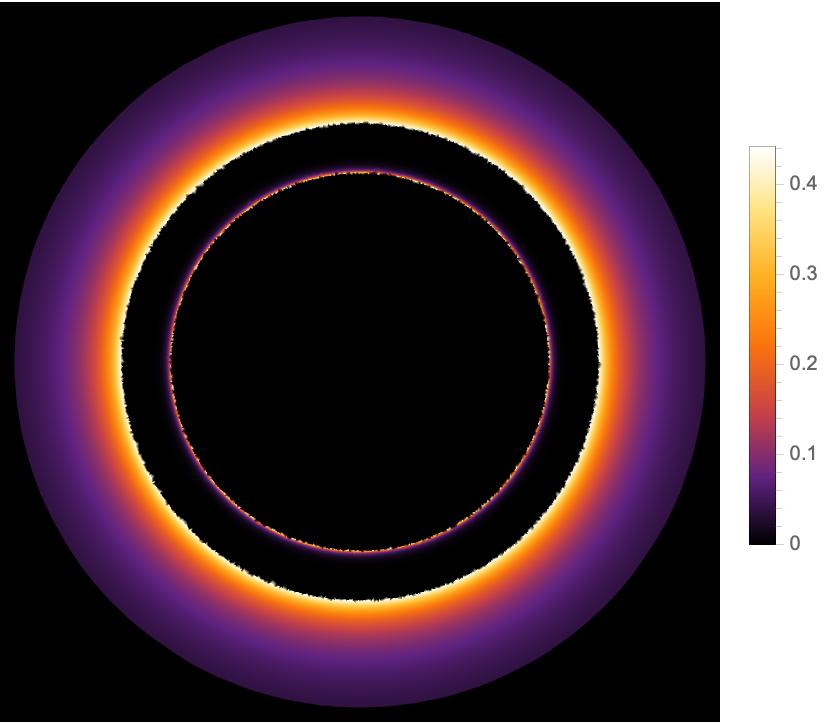}
	\hfill
	\includegraphics[width=.32\textwidth]{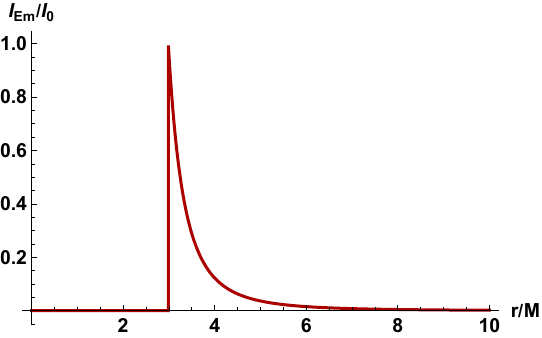}     
	\hfill
	\includegraphics[width=.32\textwidth]{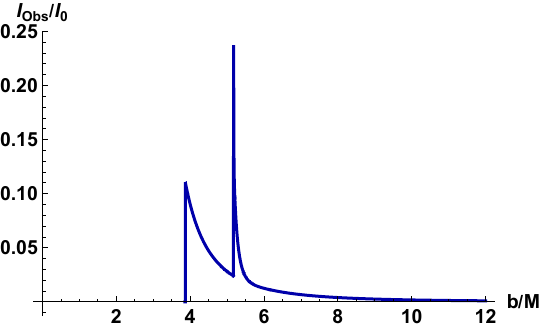}
	\hfill
	\includegraphics[width=.32\textwidth]{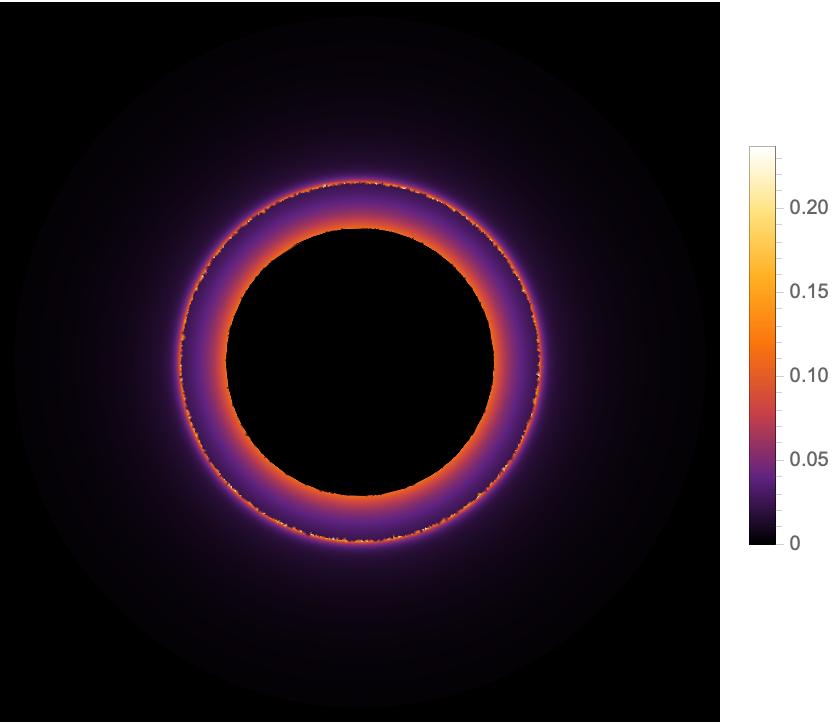}
	\hfill
	\includegraphics[width=.32\textwidth]{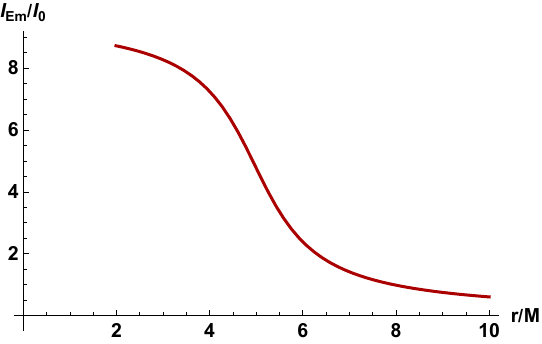}   
	\hfill
	\includegraphics[width=.32\textwidth]{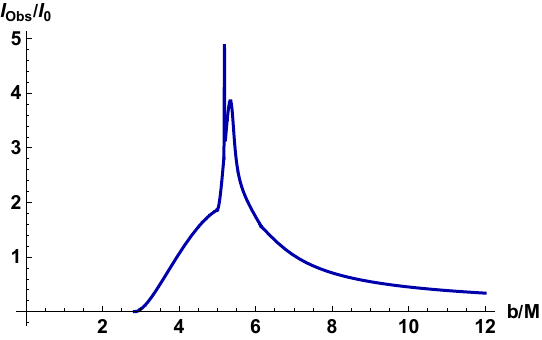}
	\hfill
	\includegraphics[width=.32\textwidth]{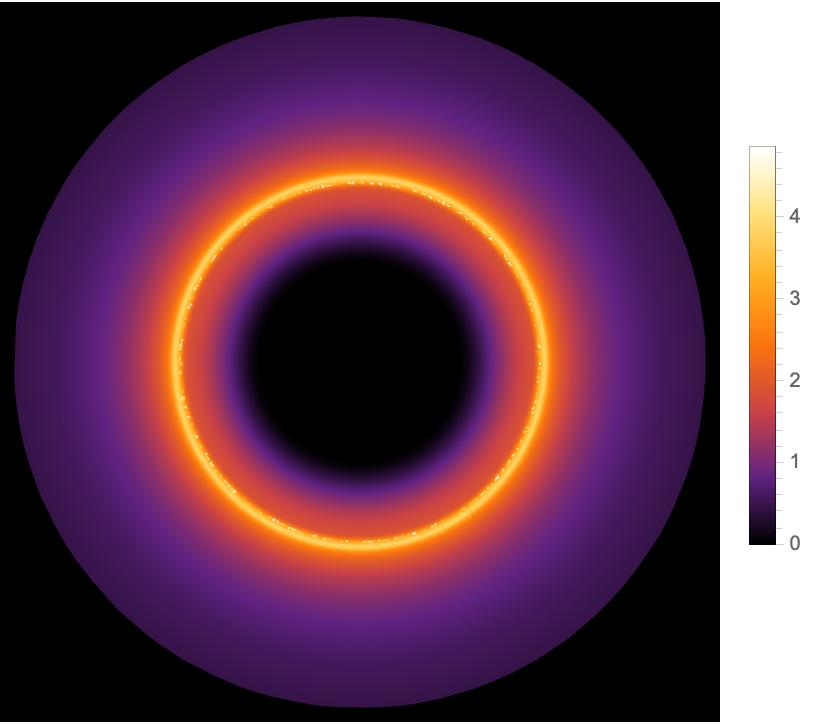}
	\caption{\label{fig4-4} 
		The overall emission intensities $I_{\rm Em}$ of optical and geometrically thin accretion disks are plotted with respect to the radius $r$ in the left column. In the middle column, the overall observed specific intensities are shown with respect to the impact parameter $b$. The right column displays the optical appearances of MCRBHs with a thin accretion disk. In each row, the emission profiles are Models I, II, and III, in that order. In all models $q = 0.01$.
	}
\end{figure}

Therefore, we initially assume that $I_{\rm Em} (r)$ follows a quadratic power decay function related to the ISCO, given by
\begin{equation}\label{EmMod1}
I_{\rm Em}^{\rm I} (r) = 
\begin{cases}
 \left(\frac{1}{r-\left(r_{\text{ISCO}}-1\right)}\right)^{2}, \qquad r> r_{\text{ISCO}},\\
 0, \qquad\qquad\qquad\qquad r\le r_{\text{ISCO}}.
\end{cases}
\end{equation}
In the first row of Figure~\ref{fig4-4}, the plots illustrate the relationship between the radius and the total emitted intensity $I_{\rm Em}^{\rm I} (r)$, as well as the impact parameter and the total observed intensity $I_{\rm Obs}(r)$ associated with Eqs. \eqref{total_Iobs} and \eqref{EmMod1}, respectively. Additionally, the two-dimensional observation characteristics in celestial coordinates are depicted.
With the model parameter $q=0.01$, the resulting MCRBH ISCO radius is approximately $r_{\rm ISCO} \simeq 5.97$. As illustrated in the left panel of the first row in Figure~\ref{fig4-4}, the emission function peaks around $r_{\rm ISCO} \simeq 5.97$ indicating the radius of the ISCO as the position where radiation emission ceases.
In the middle panel of the first row in Figure~\ref{fig4-4}, the direct emission peaks around $b \simeq 6.90$. The observed lensed ring is confined to a narrow range of $\sim5.46$ to $\sim 5.97$.
Its contribution to the total observed intensity is minimal, comprising only $\sim 5.1\%$ of the total observed intensity. The photon ring, located at $b\simeq 5.19$, is an extremely narrow ring with an almost negligible contribution to the total observed intensity, accounting for only $\sim 0.2\%$. 
The MCRBH two-dimensional observation feature is displayed in celestial coordinates in the right panel of the first row in Figure~\ref{fig4-4}. The black disk boundary represents $r_{\rm ISCO}$. The lensed ring is represented by the distinctive narrow lines in the black disk, whereas the photon ring, which appears to be considerably weaker, continues to approach the interior of the BH.

Next, we explore the scenario where the accretion flow ceases radiating at the position of the photon ring. Assuming that $I_{\rm Em}(r)$ follows a third power decay function associated with the radius of the photon ring, we can represent 
\begin{equation}\label{EmMod2}
	I_{\rm Em}^{\rm II} (r) = 
	\begin{cases}
		\left(\frac{1}{r-\left(r_{\text{Ph}}-1\right)}\right)^{3}, \qquad\,\,\,\, r> r_{\text{Ph}},\\
		0, \qquad\qquad\qquad\qquad r\le r_{\text{Ph}},
	\end{cases}
\end{equation}
in which $r_{\rm Ph}$ denotes the radius of the photon ring for the MCRBH. In the second row of Figure~\ref{fig4-4}, we present $I_{\rm Em}^{\rm II}(r)$ as a function of $r$, $I^{\rm Obs}(r)$ with respect to $b$, and a two-dimensional BH image. The left panel of the second row in Figure~\ref{fig4-4} illustrates the emission peaking at the photon ring, approximately $r_{\rm Ph} \simeq 2.99$ for $q=0.01$. As can be seen from the middle panel of the second row in Figure~\ref{fig4-4}, the observed direct emission peaked at $b \simeq 3.87$. The emission of lensed rings is bounded in the interval $5.17 \sim 5.57$. The lensed ring encompasses the photon ring, hardly visible at $b \simeq 5.17$, which is nearly indistinguishable from the lensing ring. The contribution of the lensed ring emission to the total observed intensity is $4\%$, while the photon ring continues to make an entirely negligible contribution, which is barely visible. 
The right panel of the second row in Figure~\ref{fig4-4} displays the two-dimensional observation features of the MCRBH in this scenario. It is evident that the image exhibits a distinct bright ring, indicating that the ring seen in the image contains the photon ring. 

Finally, we consider the case where the accretion flow ceases radiating at the MCRBH event horizon $r_{+}$ and goes off more gradually to zero than in the first two scenarios; in this scenario, $I_{\rm Em}(r)$ can be given as
\begin{equation}\label{EmMod3}
	I_{\rm Em}^{\rm III} (r) = 
	\begin{cases}
		\frac{\frac{\pi}{2}-\tan^{-1}\left(r-\left(r_{\text{ISCO}}-1\right)\right)}{\frac{\pi}{2}-\tan^{-1}\left(r_{\rm Ph}\right)}, \qquad r> r_{+},\\
		0, \qquad\qquad\qquad\qquad\qquad\, r\le r_{+}.
	\end{cases}
\end{equation}
In Figure~\ref{fig4-4}, the third row displays the aggregate emitted intensity function (left panel), the aggregate observed intensity function (middle panel), and the two-dimensional image (right panel) for this model. In Figure~\ref{fig4-4}, the left panel of the third row indicates that the emission reaches its highest point near the radius of the event horizon of the MCRBH, which is approximately $r_{+} \simeq 1.99$. 
The middle panel of the third row in Figure~\ref{fig4-4} displays a distinct and narrow spike at a value of $b \approx 5.18$, which corresponds to the photon ring. Additionally, there is a wider bump at $b \approx 5.34$, which represents the lensing ring. Once again, the photon ring and lensed ring are superimposed onto the direct image. 
In Figure~\ref{fig4-4}, the right panel of the third row illustrates that the optical appearance exhibits a narrow, yet distinctively brighter, extended ring. This ring is formed by the combined contributions of the direct, lensed, and photon ring emissions. However, it is generally safe to disregard the photon ring emission, since it remains entirely insignificant. Although the description provided in Figure~\ref{fig4-4} only considered a few highly idealized scenarios of thin accretion material surrounding the MCRBH, our analysis indicates that the optical appearance is primarily governed by direct emission. The contribution of lensed ring emission to the total intensity is minimal, and the photon ring influence is disregarded in all cases.
\begin{figure}[htb]
	\centering 
	\includegraphics[width=.32\textwidth]{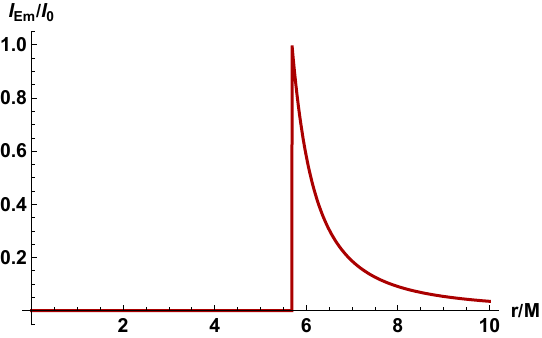}     
	\hfill
	\includegraphics[width=.32\textwidth]{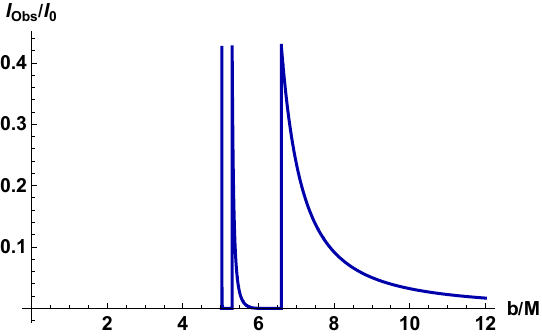}
	\hfill
	\includegraphics[width=.32\textwidth]{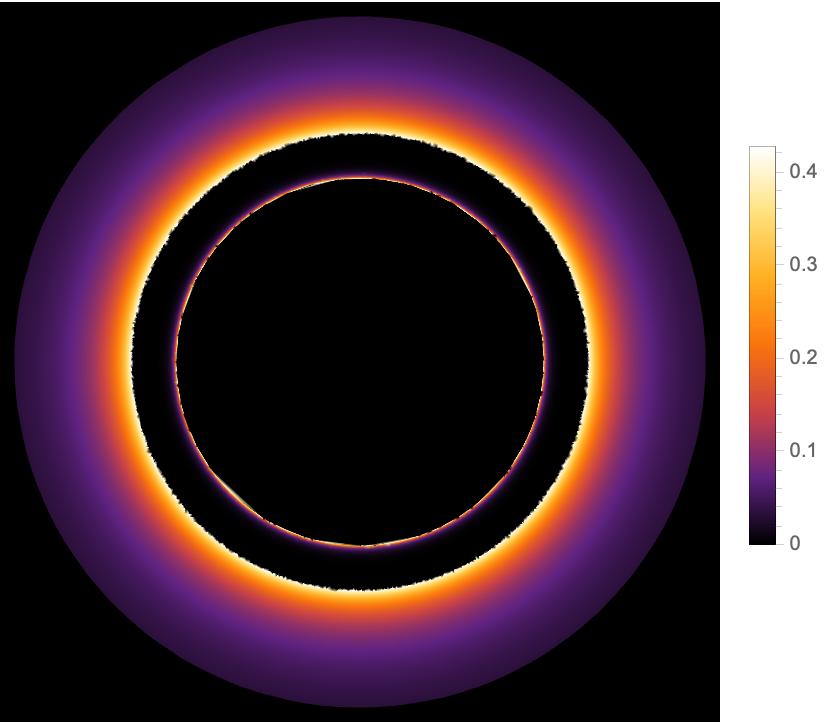}
	\hfill
	\includegraphics[width=.32\textwidth]{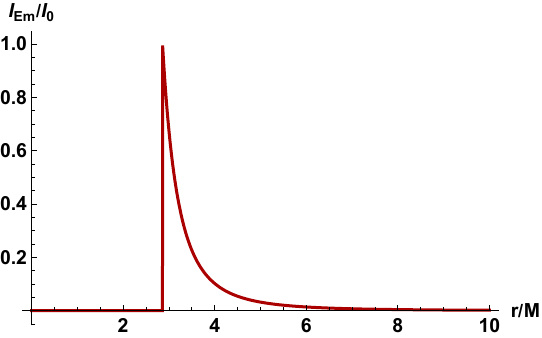}     
	\hfill
	\includegraphics[width=.32\textwidth]{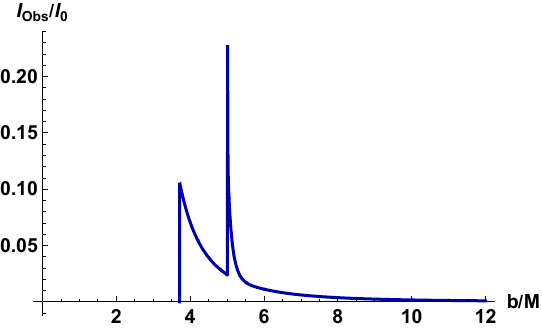}
	\hfill
	\includegraphics[width=.32\textwidth]{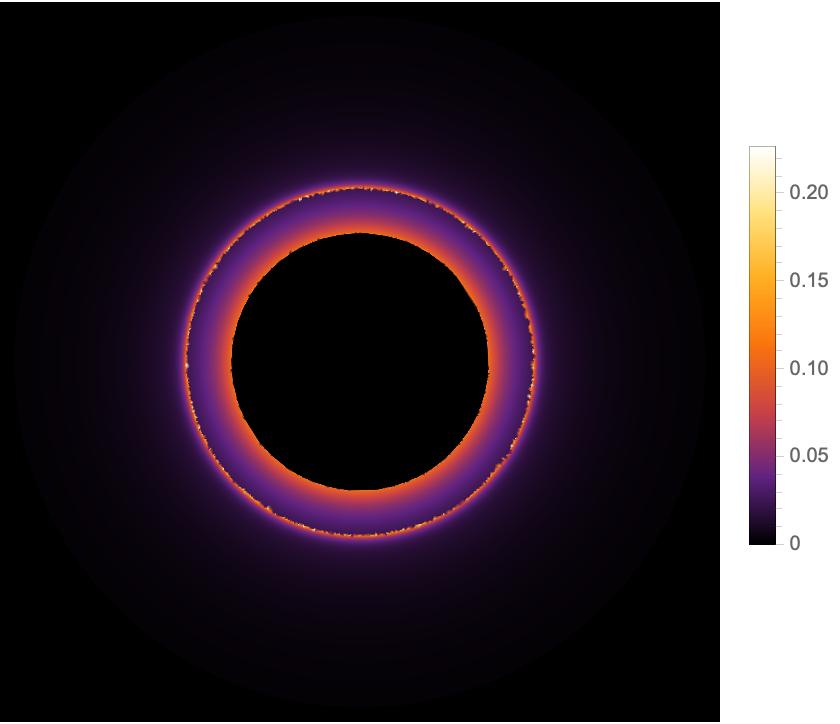}
	\hfill
	\includegraphics[width=.32\textwidth]{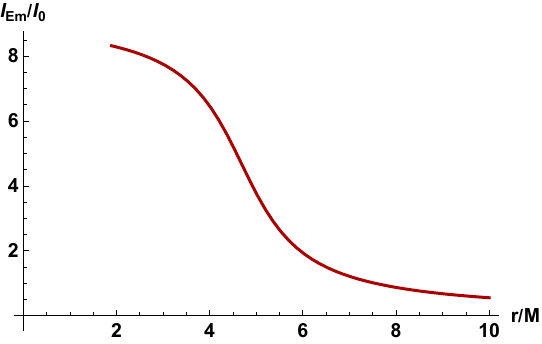}     
	\hfill
	\includegraphics[width=.32\textwidth]{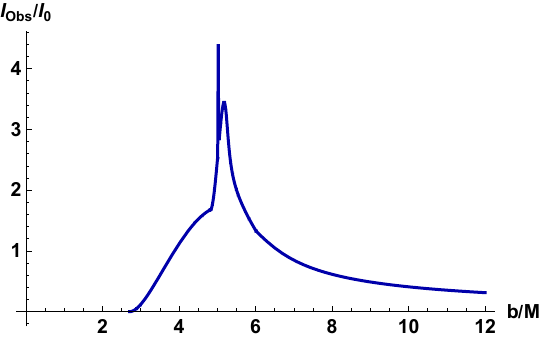}
	\hfill
	\includegraphics[width=.32\textwidth]{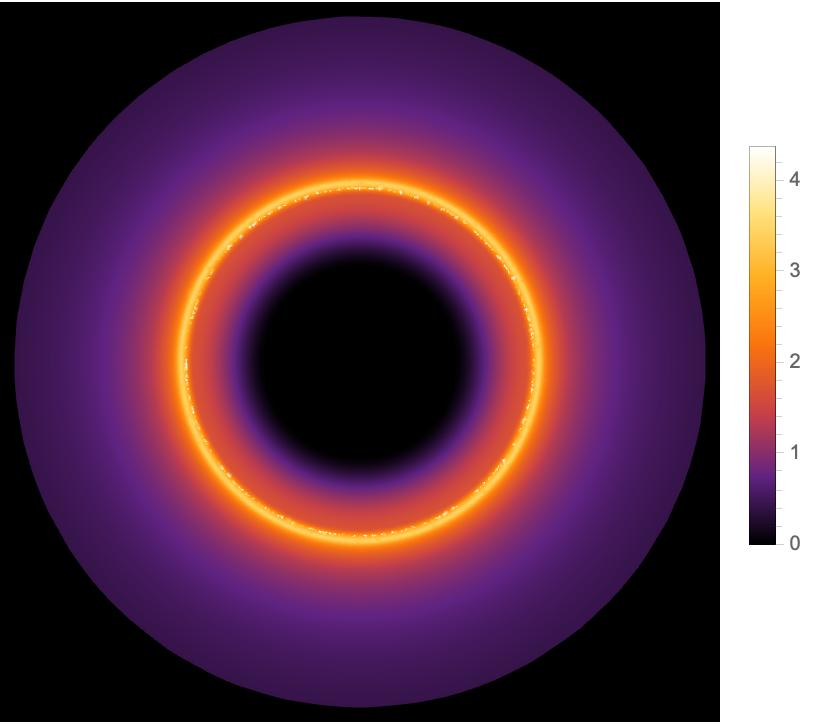}
	\caption{\label{fig4-5} 
		Here, the plot is analogous to Fig. \ref{fig4-4}, but with $q$ set to $0.1$.
	}
\end{figure}
For the case with $q=0.1$, we illustrate the emission profiles and observed intensities in Figure~\ref{fig4-5}, which exhibit similar behavior to those with $q=0.01$. However, there are differences in the intensities and the positions of the photon ring and lensed ring. From the numerical analysis of Figures~\ref{fig4-4} and \ref{fig4-5}, we found that increasing $q$ increased the thickness of the lensed and photon rings.
Notably, the BH exhibits higher intensities as the MCRBH parameter $q$ decreases.

The accretion disk flow around a BH typically arises when cosmic matter becomes captured by the BH's gravitational field and rotates with significant angular momentum. The BH is illuminated by the light rays emitted by these gas matter.
In the case of minimal angular momentum, matter will radially flow toward the BH, forming spherically symmetric accretion disk \cite{WangPRD2023,YuanARAA2014}.
In the following two subsections, we will go over the photon rings and shadows of an MCRBH with both static and infalling spherical accretions.

\subsubsection{Static spherical accretion flow}

\begin{figure}[htb]
	\centering 
	\includegraphics[width=.6\textwidth]{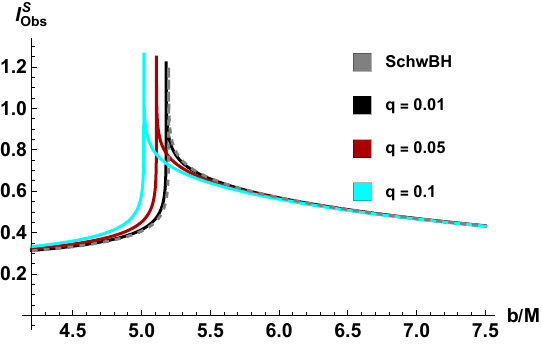}      
	\caption{\label{fig4-6}
		Plot of the total observed intensity $I^{\rm S}_{\rm Obs}$ with respect to impact parameter $b$ for the MCRBH surrounded by the static spherical accretion flow for various values of $q$. The values incorporate the Schwarzschild case, represented by a gray dashed line.
	}
\end{figure}

Let us now examine a static spherical accretion flow, which is both optically and geometrically thin, and remains statically distributed beyond the horizon of the MCRBH. 
Thus, to determine the specific intensity $I(\nu_{o})$ emitted by the accretion flow and observed by an observer at $r=\infty$ (in units of ${\rm erg}\,\, {\rm s}^{-1}\,\, {\rm cm}^{-2}\,\, {\rm str}^{-1}\,\, {\rm Hz}^{-1}$), one can integrate the specific emissivity along the photon path $\gamma$ as \cite{ZengEPJC2020,GuoPRD2022,LiEPJC2021,WaliaJCAP2023,HuEPJC2022}
\begin{equation}\label{SpeInten}
I(\nu_{o}) = \int_{\gamma} g^{3}j_{e}\left(\nu_{e}\right)dl_{\rm prop}.
\end{equation}
Here, $g = \nu_{o}/\nu_{e}$ represents the redshift factor, where $\nu_{o}$ and $\nu_{e}$ denote the observed and emitted photon frequencies, respectively. $j_{e}(\nu_{e})$ stands for the emissivity per unit volume in the rest frame, and we take $j_{e}(\nu_{e}) \propto \delta(\nu_{r}-\nu_{r})/r^{2} $, with $\nu_{r}$ being the emitter's rest-frame frequency. Besides, $dl_{\rm prop}$ signifies the infinitesimal proper length.
In this scenario, the redshift factor $g$ and the infinitesimal proper length $dl_{\rm prop}$ are given  by
\begin{equation}\label{InfPropLength}
g = \sqrt{\text{A}(r)}, \qquad dl_{\rm prop} = \sqrt{\frac{1}{\text{A}(r)}+r^{2}\left(\frac{d \varphi}{dr}\right)^{2}}\, dr,
\end{equation}
respectively, where the inverse of Eq. \eqref{veff} yields $d\varphi/dr$.
Therefore, using Eqs. \eqref{SpeInten} and \eqref{InfPropLength} along with the associated assumptions, the specific intensity observed by a static observer at infinity can be found as:
\begin{equation}\label{Isobs}
I_{\rm Obs}^{\rm S}(r) = \int_{\gamma} \frac{{\rm A}^{3/2}(r)}{r^{2}}\sqrt{\frac{1}{{\rm A}(r)}+\frac{b^{2}}{r^{2}-b^{2}{\rm A}(r)}}\,dr.
\end{equation}
The observed specific intensity $I_{\rm Obs}^{\rm S}$ varies with the impact parameter $b$ and is influenced by the parameter $q$. Figure~\ref{fig4-6} illustrates the $I_{\rm Obs}^{\rm S}$ emitted by a static spherical accretion flow around the MCRBH. Meanwhile, spacetime symmetry is expected for $I_{\rm Obs}^{\rm S}$ corresponding to negative $b$. In the positive $b$ region, regardless of the value of $q$, the specific intensity $I_{\rm Obs}^{\rm S}$ increases with $b$ and rapidly peaks at the critical impact parameter $b_{\rm C}$, then gradually decreases to a minimum value as $b$ increases.
As the parameter $q$, associated with the magnetic charge of the BH, rises, the intensity also increases. This implies that the luminosity around the MCRBH surpasses that of the Schwarzschild scenario. Such evidence enables us to distinguish between the two cases.
Furthermore, it is apparent that a higher value of $q$ correlates with a reduced shadow radius.

\begin{figure}[htb]
	\centering 
	\includegraphics[width=.312\textwidth]{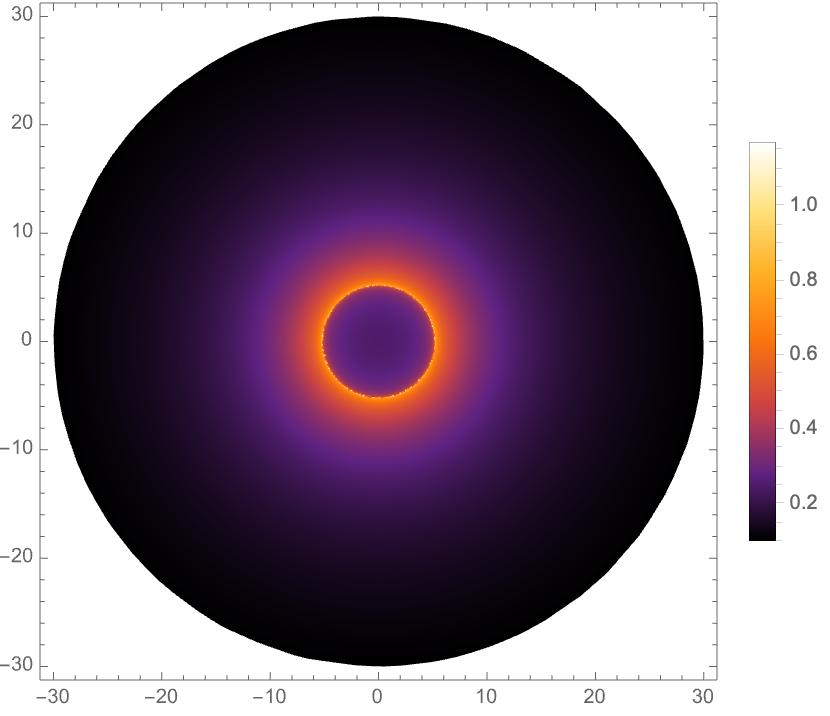}       
	\hfill
	\includegraphics[width=.312\textwidth]{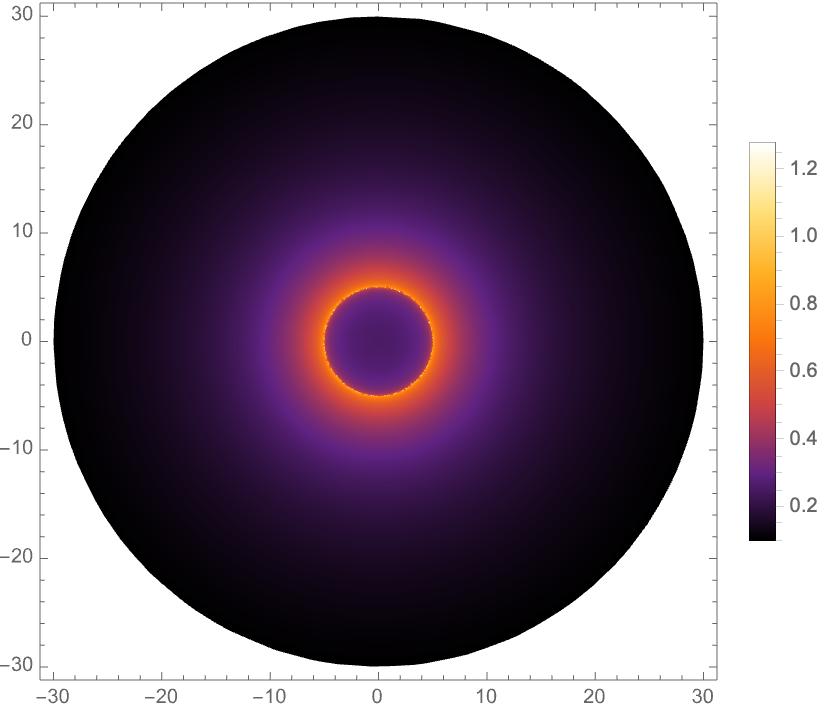}
	\hfill
	\includegraphics[width=.35\textwidth]{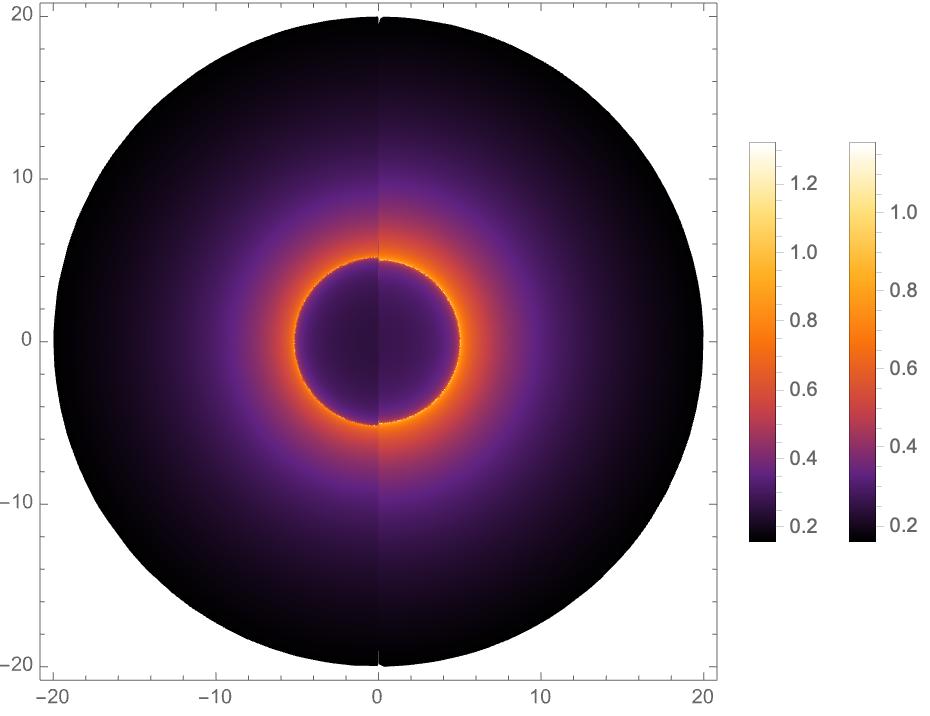}
	\caption{\label{fig4-7}
Plot of the two-dimensional shadow images and photon rings of MCRBHs surrounded by a static spherical accretion flow.  
The parameter $q$ takes values of $0.01$ and $0.1$ for the left and middle panels, respectively. Increasing $q$ decreases the shadow radius but enhances the slightly luminosity of both shadows and photon rings.
The right panel, representing a magnified view of the left and middle panels, showcases the distinction in shadow radii and luminosities between $q=0.01$ and $q=0.1$ cases, respectively.
}
\end{figure}

As illustrated in Figure~\ref{fig4-7}, the observed intensity distribution \eqref{Isobs} in a two-dimensional plane should indicate the MCRBH illuminated by static spherical accretion for equatorial observers. A ``shadow'' refers to the dark spot within a luminous ring. Given that some of the radiation from the accretion flow within the photon ring may escape to infinity, the shadow is not completely dark and does not have zero intensity. 
The shadows and photon rings of the MCRBH exhibit higher luminosities compared to those of the Schwarzschild BH. The parameter $q$ associated with the MCRBH contributes to reducing the curvature of spacetime, allowing more photons to escape the event horizon of the BH.
Therefore, as the parameter $q$ increases, the shadow radius shrinks along with a smaller photon ring. This finding aligns with the results shown in Figures~\ref{fig4-6} and \ref{fig4-7}, as well as the trend of increasing specific intensity with $q$.

\subsubsection{Infalling spherical accretion flow}
\begin{figure}[htb]
	\centering 
	\includegraphics[width=.6\textwidth]{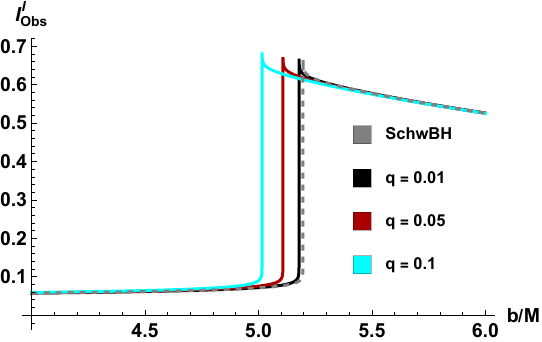}        
	\caption{\label{fig4-8}
		Plot of the total observed intensity $I^{\rm I}_{\rm Obs}$ with respect to impact parameters $b$ for the MCRBH surrounded by the infalling spherical accretion flow for various values of $q$. The values incorporate the Schwarzschild case, represented by a gray dashed line.
	}
\end{figure}

\begin{figure}[htb]
	\centering 
	\includegraphics[width=.312\textwidth]{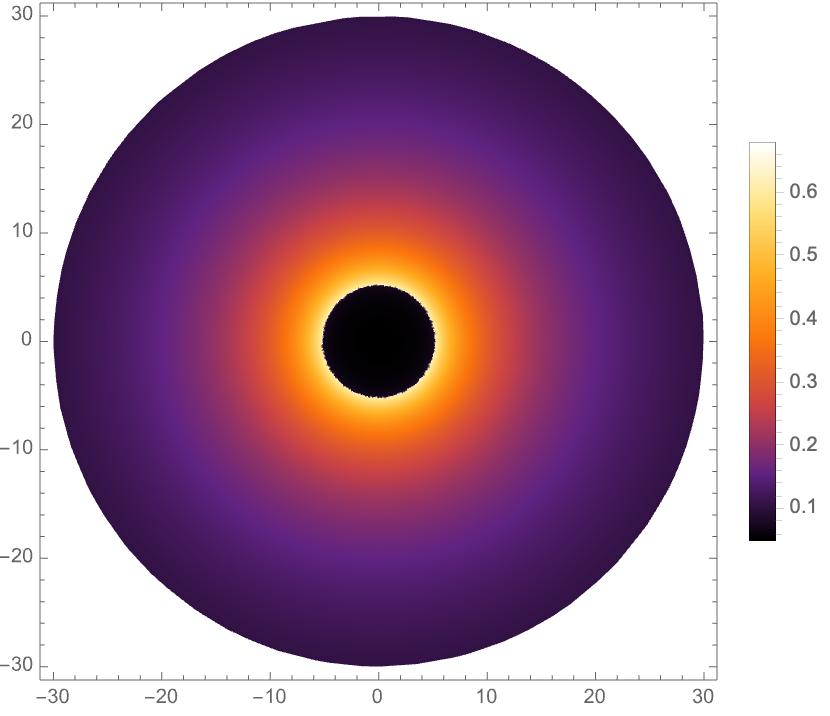}       
	\hfill
	\includegraphics[width=.312\textwidth]{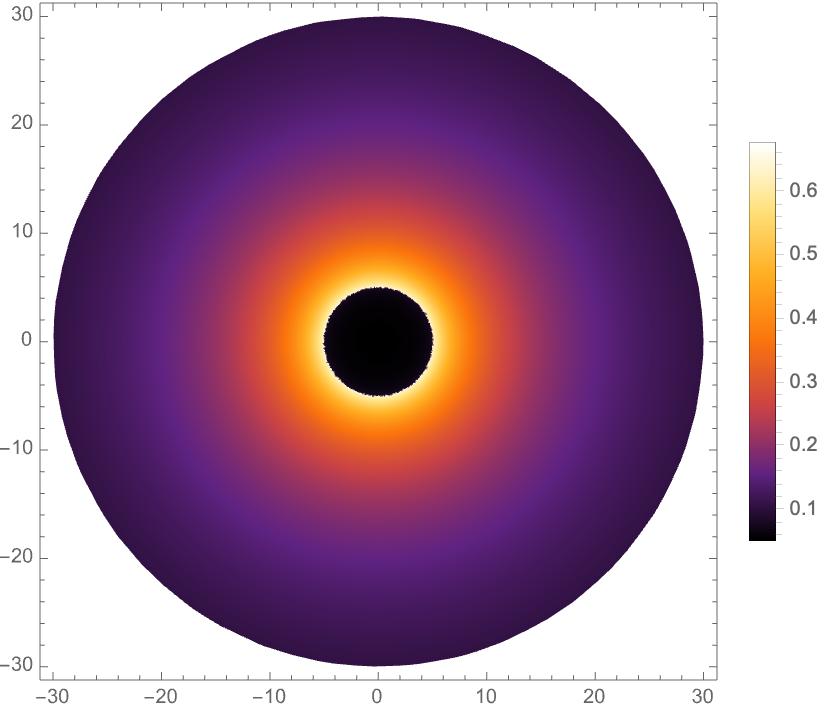}
	\hfill
	\includegraphics[width=.35\textwidth]{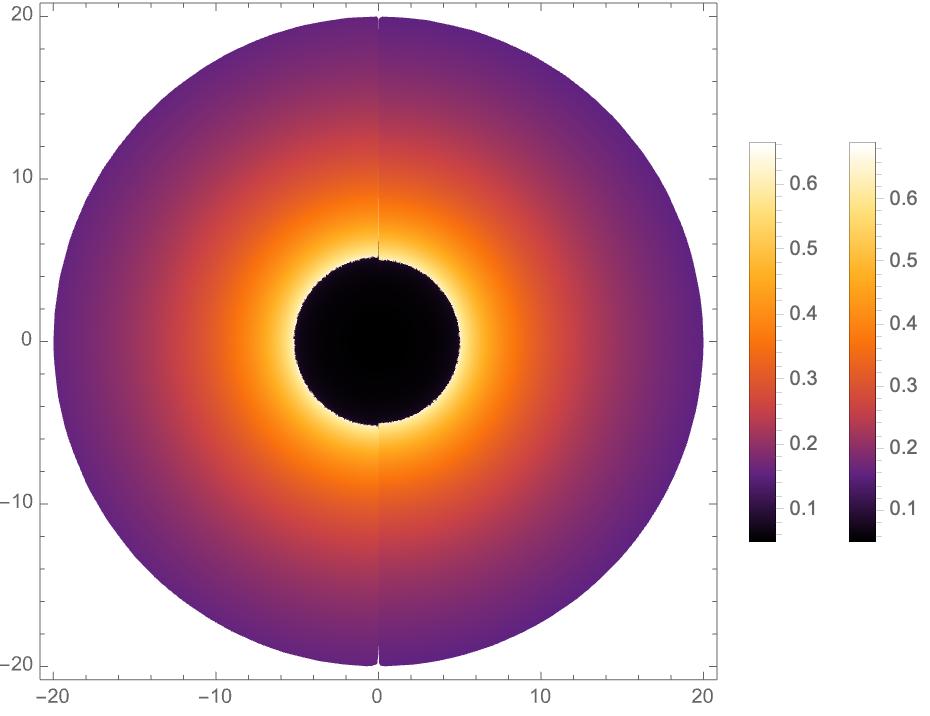}
	\caption{\label{fig4-9}
		Analogous to Fig. \ref{fig4-7}, only with an infalling spherical accretion flow instead of a static one.
	}
\end{figure}

In this subsection, we will explore a more realistic scenario where the MCRBH is enveloped by a radially infalling spherical accretion disk, reflecting the dynamic nature of matter in the universe. In this context, the redshift factor deviates from that of the static spherical accretion flow, as follows \cite{ZengEPJC2020,ZengEPJC2020-2,WangPRD2023,GuoPRD2022,WaliaJCAP2023,HuEPJC2022,ShaikhMNRAS2019}
\begin{equation}\label{RedShiftInfalling}
g = \frac{k_{\mu}u_{o}^{\mu}}{k_{\nu}u_{e}^{\nu}},
\end{equation}
where $k^{\beta} = \dot{x}_{\beta} = \partial x^{\beta}/ \partial \lambda$ 
denotes the four-velocity of the photon emitted from accretion matter, where $\lambda$ represents the affine parameter \cite{LiEPJC2021,WangPRD2023}, providing us with
\begin{equation}\label{4VelocityPhoton}
k_{\beta} = \left(\frac{1}{b},\pm \frac{1}{{\rm A}(r)b}\sqrt{1-\frac{{\rm A}(r)}{r^{2}}b^{2}}\,,0,\pm 1\right),
\end{equation}
according to the null geodesic.
As photons approach or flee from the black hole, signs $\pm$ in $k_r$ indicate their radial inward or outward motion, but in $k_\varphi$, they indicate their counterclockwise and clockwise motion, respectively.
In Eq. \eqref{RedShiftInfalling}, we denote $u^{\beta}_{o} = (1, 0, 0, 0)$
as the four-velocity of the static observer at infinity, and the four-velocity of the infalling accretion, $u^{\beta}_{e}$, is then expressed as:
\begin{equation}\label{4VelocityinfallingAcc}
u^{\beta}_{e} = \left(\frac{1}{{\rm A}(r)},-\sqrt{1-{\rm A}(r)}\,,0,0\right).
\end{equation}
Thus, considering Eqs. \eqref{4VelocityPhoton} and \eqref{4VelocityinfallingAcc}, the redshift factor provided in Eq. \eqref{RedShiftInfalling} can be rewritten as follows
\begin{equation}\label{RedShiftInfalling2}
g = \left(u^{t}_{e}+\frac{k_{r}}{k_{t}}u^{r}_{e}\right)^{-1},
\end{equation}
and the proper distance can be determined as follows \cite{BambiPRD2013,ShaikhMNRAS2019}
\begin{equation}\label{PropDist}
dl_{\rm prop} = k_{\mu} \mu^{\mu}_{e} d\lambda = \frac{k_{t}}{g|k^{r}|}dr .
\end{equation}
In a simplified model, we also take into account that the specific emissivity is monochromatic, thus the emissivity $j_{e}\left(\nu_{e}\right)$ remains identical to that of static spherical accretion. 
Thus, integrating the observed specific intensity, given in Eq. \eqref{SpeInten},  over all observed photon frequencies, the total observed intensity of the MCRBH with a radially infalling spherical accretion flow can be expressed as:
\begin{equation}\label{Iiobs}
	I_{\rm Obs}^{\rm I}(r) \propto \int_{\gamma} \frac{g^{3} k_{t}}{r^{2}|k^{r}|}dr .
\end{equation}
As presented in Figures~\ref{fig4-8} and \ref{fig4-9}, analogous to the static accretion scenario, we investigate the shadow image and luminosity distribution of the MCRBH, which is encompassed by the infalling spherical accretion, applying Eq. \eqref{Iiobs}. 
In this scenario, we begin by illustrating the impact of the parameter $q$ on the specific intensity $I_{\rm Obs}^{\rm I}$ of the MCRBH.
As depicted in Figure~\ref{fig4-8}, we note a sharp rise in the specific intensity $I_{\rm Obs}^{\rm I}$ with increasing impact parameter $b_{\rm C}$, reaching its peak at $b = b_{\rm C}$. This behavior holds true for various values of $q$ responsible for the BH magnetic charge parameter.
In the region where $b > b_{\rm C}$, the specific intensity $I_{\rm Obs}^{\rm I}$ demonstrates a declining trend for a fixed parameter $q$. Moreover, as $b$ approaches infinity $(b\rightarrow \infty)$, the specific intensity $I_{\rm Obs}^{\rm I}$ will asymptotically approach zero $(I_{\rm Obs}^{\rm I}\rightarrow 0)$.
In the case of Schwarzschild (shown by the dashed gray line), the intensity is lower than that of the MCRBH. As we increase the parameter $q$ of the MCRBH, the peak value of the specific intensity $I_{\rm Obs}^{\rm I}$ increases, but it drops at the same rate. 

To explore the traits of $I_{\rm Obs}^{\rm I}$, which resemble those shown in Figure~\ref{fig4-6}, we vary the value of $q$. As such, we consider $q$ values of $0.01, 0.05$, and $0.1$, where the peak of each curve occurs at $b = b_{\rm C}$, as previously mentioned. The highest value of $I_{\rm Obs}^{\rm I}$ is determined from the light rays that marginally escape from the BH, and this value grows as $q$ increases. Furthermore, the size of the BH shadow image shrinks as the parameter $q$ increases.
Comparing the observed intensities depicted in Figures~\ref{fig4-6} and \ref{fig4-8} allows us to see the disparity between static and infalling spherical accretions. We observe that, for a given parameter $q$, the intensity of the BH image in the infalling spherical accretion scenario is lower than that in the static case, attributed to the Doppler effect.

Next, Figure~\ref{fig4-9} illustrates the MCRBH image in a two-dimensional plane for two various values of $q$ encompassed by the infalling accretion flow. Comparatively, the brightness of the central BH shadow, encircled by a luminous photon ring, is noticeably diminished compared to its static counterpart. 
In comparison to a Schwarzschild BH, the presence of $q$, which represents the magnetic charge parameter of the BH, causes the photon ring to shrink. Consequently, this results in a smaller BH shadow observed by a distant observer. When illuminated by a spherical accretion flow, the deviations in the shadow region, luminous photon ring, and luminosity distribution caused by the $q$ parameter of the MCRBH diverge significantly from those of a Schwarzschild BH.

\section{Discussion and conclusions}\label{SecC}
The scarcity of magnetically charged regular regular black holes (MCRBH), characterized by their mass and a model parameter $q$, from the coupling of Einstein gravity and NLED satisfying to Maxwell’s weak field limit,  hinders the expansion of NLED tests with lower charge through observations, like those from the EHT. 
Furthermore, the influence of the lower model parameter on the astrophysical environment of MCRBHs remains an area requiring further investigation.
By considering the spacetime structure, null, and time-like geodesics around the MCRBH, we study scalar invariants, circular orbits, and shadow silhouettes,
as well as modeling the MCRBHs as supermassive BHs M87* and Sgr A*, using the EHT results to constrain $q$, which has impacts close to the RN scenario in the region that is asymptotically flat. 
Our investigation is then extended to various aspects of the accretion disk around the MCRBH, employing the thin-disk approximation. To do this, we explore the influence of $q$ on the physical characteristics of the thin accretion disk  through the Novikov-Thorne model, and probe the observed shadow images, rings, and optical appearance of the MCRBH with both, thin disk-shaped and spherical accretion models. In doing so, our findings reveal distinct traces of the model parameter $q$ in MCRBH physics as follows:
\begin{enumerate}[(I)]
	\item 
	We first explored the horizon structure of the MCRBH, finding that it expands as the parameter $q$ decreases.
	We then discussed the scalar invariants, namely the Ricci and Kretschmann scalars, for the MCRBH metric and investigated the spacetime structure near it. Our findings indicate that these scalars are well-defined at the BH center, suggesting the absence of a curvature singularity. Moreover, as $q$ increases, the scalar invariants at the BH origin decrease.
	\item 
	We conducted an analysis of circular orbits regarding the behavior of both time-like and null geodesics. It was found that stable circular orbits are situated at large distances, whereas unstable circular orbits are situated at small radii.
	We observed that for larger values of $q$, the characteristic radii $(r_{-}, r_{+}, r_{\rm Ph}, b_{\rm C}, r_{\rm MOB}, r_{\rm ISCO})$ decreased. It was also noted that for a distant observer from a BH with an asymptotically flat metric, the shadow radius reduced to the critical impact parameter. 
	Due to the strong influence of the parameter $q$ near the MCRBH, it significantly affected the behavior of the photon sphere. In turn, we found that the shadow size decreased with higher values of $q$,  where $q$ is consistent with Maxwell’s weak field limit.
	Nonetheless, the deviations in the corresponding shadow radius for observers at infinity approached the RN case. Furthermore, constraints were put on $q$ based on the bounds inferred by the EHT on the Schwarzschild shadow radius of M87* and Sgr A*. A lower range for $q$ was found in Sgr A*, while a higher one was found in M87* at the $1\sigma$ confidence level. On the other hand, the results for Sgr A* imposed more robust constraints on $q$ than those for M87*. Therefore, the EHT observations do not exclude the potential MCRBHs at galactic centers in a consistent finite parameter space of the MCRBHs.
	Next, we discussed the behavior of the specific energy, specific angular momentum, and angular velocity of particles in circular motion within the equatorial plane, and it was observed that as $q$ increased, these quantities decreased in the radial profile.
	\item 
	We then explored the impact of $q$ on the time-averaged energy flux, differential luminosity, and disk temperature generated by the thin accretion disk in the equatorial plane around the MCRBH.
	We observed that an increase in the parameter $q$ of the MCRBH leads to an increase in all the aforementioned physical quantities. Besides, the peak values shift closer to the interior boundary of the disk, and there is a slight deviation from the radiation spectrum of a Schwarzschild BH. Likewise, the disk is brighter and warmer than the disk surrounding a Schwarzschild BH $(q \rightarrow 0)$ in GR.
	\item 
	By considering the MCRBH surrounded by an optically and geometrically thin disk accretion, we probed and categorized the light trajectories near the corresponding BH using the Gralla-Holz-Wald criteria \cite{GrallaPRD2019}, which depend on the number of their intersections with the equatorial plane. 
	As such, we observed that the outer region of the shadow contains the photon rings and lens ring, respectively, in addition to the dark central shadow region, as seen by an observer. 
	We obtained the range of these three types of ray trajectories and plotted them using the ray tracing code. We found that the width of the lensed and photon rings increased with the increase of $q$. Also as $q$ increases, these two light rings become thicker.
	Then, three toy models of emission profiles were taken as examples to explore how the brightness contributed to the total observed intensities from the direct, lensed ring, and photon ring intensities.
	We also revealed that, although the photon ring accumulated more than three times through the thin disk accretion and acquired brightness, it remained invisible to the observer due to its extreme demagnetization.
	Due to the significant demagnetization, the lensed ring only contributes a minor proportion to the overall observed specific intensity. The primary contribution of the observed specific intensity is associated with the direct emissions.
	\item 
	We finally assumed that a static and infalling spherical accretions flow, respectively, illuminates the MCRBH. 
	In both spherical accretion models, a luminous photon ring significantly surrounded a dark region, indicating the BH shadow. Compared to the Schwarzschild BH, the shadow size for the MCRBH was lower, while the photon ring was more brilliant. The former was distinguished from the latter by this significant feature.
	When the model parameter $q$ is fixed, the shadow size remains the same in both accretion scenarios. However, the luminosity of the infalling accretion is dimmer than the static accretion due to the Doppler effect. 

\end{enumerate}

Ultimately, our theoretical investigations have been quite idealized and involved accretion models that are, in principle, far from realistic astrophysical settings; nonetheless, this study presented a potential avenue for distinguishing MCRBHs from static BHs in GR.	
We expect that our current findings could contribute to future studies testing Maxwell's theorem using the optical appearance of BHs.

\acknowledgments
The research of L.M.N. and S.Z. was supported by 
the European Union.-Next Generation UE/MICIU/Plan de Recuperacion, Transformacion y Resiliencia/Junta de Castilla y Leon, 
RED2022-134301-T financed by MICIU/AEI/10.13039/501100011033, 
and PID2020-113406GB-I00 financed by MICIU/AEI/10.13039/501100011033.
The research of S.H. Dong was partially supported by the project 20240220-SIP-IPN and started this work on the research stay in China.
The research of X.H.F. was supported by NSFC (National Natural Science Foundation of China) Grant No. 11905157 and No. 11935009.

\end{document}